\documentclass[pdflatex,sn-mathphys-num]{sn-jnl}
\usepackage{graphicx}
\usepackage{multirow}
\usepackage{amsmath,amssymb,amsfonts}
\usepackage{booktabs}
\usepackage{xcolor}
\usepackage{float}
\usepackage{array}
\usepackage{enumitem}
\usepackage{url}
\usepackage{tikz}
\usepackage[normalem]{ulem} % eli added
\newcommand{\etal}{\textit{et al.}}
\newcommand{\km}{$k$-means}
\newcommand{\kmc}{\km\ clustering}
\usetikzlibrary{trees,positioning}
\graphicspath{{figures/}}
\begin{document}

\title{Progress and Prospect of AI in ARPES Workflow}

\author*[1,2]{\fnm{Sandy Adhitia} \sur{Ekahana}}\email{sekahana@andrew.cmu.edu}\email{sekahana@lbl.gov}\equalcont{These authors contributed equally to this work.}
\author[1]{\fnm{Aalok} \sur{Tiwari}}\equalcont{These authors contributed equally to this work.}
\author[1]{\fnm{Pratik} \sur{Saud}}
\author[2]{\fnm{Aaron} \sur{Bostwick}}
\author[2]{\fnm{Chris} \sur{Jozwiak}}
\author[2]{\fnm{Eli} \sur{Rotenberg}}
\author*[1]{\fnm{Jyoti} \sur{Katoch}}\email{jkatoch@andrew.cmu.edu}

\affil[1]{\orgdiv{Department of Physics}, \orgname{Carnegie Mellon University}, \orgaddress{\city{Pittsburgh}, \postcode{15213}, \state{PA}, \country{USA}}}
\affil[2]{\orgname{MAESTRO, Advanced Light Source, Lawrence Berkeley National Laboratory}, \orgaddress{\city{Berkeley}, \state{CA}, \country{USA}}}

\abstract{
Artificial intelligence (AI) is becoming an increasingly useful tool across the experimental sciences, including angle-resolved photoemission spectroscopy (ARPES), which routinely produces large, multidimensional datasets of electronic structure. Recent advances in AI and machine learning (ML) have opened new opportunities  across the entire ARPES workflow, from automated sample preparation and real-time data  acquisition to post-experiment data analysis and comparison with theoretical calculations. Despite this progress, a comprehensive review of ML  applications, their capabilities, and reliability across the different stages of ARPES workflow is still lacking. In this review, we first introduce  ML methods that are most relevant to experimentalists working in condensed matter physics and materials science. We then follow the ARPES workflow, reviewing existing ML applications at each step and discussing their advantages, limitations and potential for future development. We also examine the current ARPES data landscape, where several open databases are available but remain relatively small and fragmented compared with large, shared datasets such as ImageNet. Given these limitations, we suggest that the community focus on sharing pretrained models that can be further trained, adapted to specific tasks, and redistributed, while working toward a larger and standardized open ARPES dataset repository. Finally, we discuss our perspectives on the future of AI within the ARPES workflow using a six-level framework of laboratory automation, highlighting the opportunities and challenges in moving toward a fully autonomous, self-driving ARPES laboratory.
}


\maketitle

%% ====================================================================

\section{Introduction}
\label{sec:intro}

The rapid advancement of deep neural networks following AlphaGo's 2016 milestone~\cite{silver2016mastering}, combined with widely accessible Python frameworks such as TensorFlow, PyTorch, and scikit-learn~\cite{abadi2016tensorflow,paszke2019pytorch,pedregosa2011scikit}, has made artificial intelligence (AI) a practical and powerful tool for both experimental and theoretical scientific research. Although the conceptual foundations of neural computation date back to 1943~\cite{mcculloch1943logical}, with early neural-network applications emerging in experimental particle physics in the late 1980s~\cite{denby1988}, sustained progress was enabled by the dramatic increase in computational power brought about by GPU-based computing in the 2010s. AI for science now spans diverse areas, including protein structure prediction~\cite{jumper2021highly}, global weather forecasting~\cite{lam2023learning}, fusion-plasma control~\cite{degrave2022magnetic}, and antibiotic discovery~\cite{stokes2020deep}, as highlighted in broader reviews of machine learning in physical sciences~\cite{wang2023scientific,carleo2019}.

Within condensed matter physics, systematic machine learning (ML) of the wavefunction has been shown to reduce the complexity of strongly correlated systems to a tractable computational form. Carleo and Troyer~\cite{carleo2017} introduced a variational representation of quantum states based on artificial neural networks with a variable number of hidden neurons. Their variational Monte Carlo scheme efficiently captures the complexity of entangled many body systems at low dimensions, providing accurate dynamics of prototypical interacting spin models. In addition, neural networks are trained to encode information about ordered phases by learning the order parameter, without any access to the Hamiltonian~\cite{carrasquilla2017}, and related schemes classify quantum phases in a model independent fashion~\cite{mahlow2023model}. Moving from ordered phases to bulk materials, faithful crystal-structure and symmetry representations allow crystal graph and related neural networks to predict topological, magnetic, and energetic properties~\cite{nop2025faithful}. Related sparse-representation approaches similarly accelerate estimation of 2D material properties from lattice and defect configurations~\cite{kazeev2023sparse}.

On the experimental side, analytical tools are increasingly being integrated directly into instrument workflows, enabling real-time data processing and analysis rather than relying solely on offline post-processing. For instance, machine-learning approaches for neutron scattering can provide real-time feedback to guide experiments, while simultaneously extracting material parameters and constructing phase diagrams using nonlinear autoencoders~\cite{samarakoon2022integration}. Similar real-time paradigms are rapidly expanding across experimental facilities, including ML-guided scanning transmission electron microscopy that dynamically steers measurements~\cite{kalinin2023}, synchrotron closed-loop feedback for autonomous decision making and real-time monitoring~\cite{pithan2023}, and implicit neural representations for extraction and refinement of physical parameters~\cite{chitturi2023}.

Within the fields of scattering and spectroscopy, Chen~\emph{et~al.} survey ML applications across neutron scattering, X-ray absorption, X-ray scattering, and photoemission~\cite{chen2021review}, highlighting how modern neural network architectures not only accelerate the extraction of structural and dynamical information from complex datasets, but are also increasingly integrated directly into experimental workflows. For ARPES specifically, Pustovit and Lytveniuk provide a more focused review of the convolutional-neural network approaches for ARPES data processing, primarily addressing denoising and dispersion extraction~\cite{pustovit2024}.

Here we provide a comprehensive review of AI and ML across the complete ARPES workflow. We first introduce the AI terminology relevant to experimentalists and then follow the ARPES pipeline from sample growth \& preparation, through motorized \& automated data acquisition, to real-time \& post-experiment analysis of band dispersions and self-energies, as well as ML-based surrogates for density functional theory (DFT) and GW calculations for theoretical interpretation of experimental data. We further examine the current ARPES data landscape, characterized by sporadic open repositories due to the absence of a centralized ARPES database. Finally, we discuss how standardized datasets and the circulation of pretrained models could provide a foundation for increasingly autonomous and ultimately fully automated ARPES experiments.

\section{Angle-Resolved Photoemission Spectroscopy}
\label{sec:arpes-basics}

ARPES is a powerful tool to visualize the occupied electronic band structure of crystalline materials~\cite{damascelli2003,cattelan2018perspective,lv2019angle,hofmann2021,sobota2021,zhang2022angle}. However, advances in instrumentation have substantially increased the dimensionality and complexity of ARPES measurements, creating a growing need for automated and ML-assisted approaches to data acquisition, visualization, and analysis. ARPES has evolved from one-dimensional measurement of photoelectron intensity as a function of energy to two dimensional (2D) maps, $I(E_{\rm kin},\vartheta)$, acquired along the slit of a hemispherical analyzer, and subsequently to three-dimensional dataset obtained by sample rotation or electron beam deflection. Varying the photon energy, $h\nu$, provides access to the perpendicular momentum and unlocks the comprehensive electronic structure of quantum materials. Spin and time-resolved measurements further expand the accessible parameter space. 
With the advent of third-generation light sources featuring micro-focused spot sizes under 50~$\mu$m, it became possible to scan spatial \((x,y)\) coordinates, allowing for measurement of multiple domains and the underlying physics in a single sample~\cite{cattelan2018perspective,hofmann2021}.
Consequently, modern ARPES can generate high-dimensional datasets spanning real and reciprocal space, energy, spin, time, and experimentally controlled parameters like strain, applied electric, and magnetic fields. The constantly increasing scale and complexity of these multidimensional datasets make it time consuming to choose what to measure next in real-time using manual data analysis and visualization tools. AI-assisted ARPES workflow has a potential to accelerate electronic structure research feedback.

Based on the principles of the photoelectric effect, a photon of energy $h\nu$ ejects an electron from a material system whose kinetic energy satisfies
\begin{equation}
  E_{\rm kin} = h\nu - \phi - |E_{B}|,
  \label{eq:ekin}
\end{equation}
where $\phi$ is the work function and $E_{B}$ is the binding energy prior to emission. Combined with the emission angle, this measurement maps crystal momentum and thereby reconstructs Fermi surfaces and quasiparticle dispersions. Within the sudden approximation, the measured intensity $I(\mathbf{k},\omega)$  is proportional to the single particle spectral function $A(\mathbf{k},\omega)$ multiplied by a photoemission matrix element,
\begin{equation}
  I(\mathbf{k},\omega) \propto A(\mathbf{k},\omega)\,|M_{f,i,\mathbf{k}}|^2,
  \label{eq:intensity}
\end{equation}
where matrix element $|M_{f,i,\mathbf{k}}|^2$ gives the transition probability between initial and final states depending on geometry, photon kinematics, and orbital character~\cite{moser2017experimentalist}.

The spectral function $A(\mathbf{k},\omega)$ governs the overall line shape of the spectrum and carries the many-body physics through the
retarded \emph{Green's} function,
\begin{equation}
  A(\mathbf{k},\omega)=-\tfrac{1}{\pi}\,\mathrm{Im}\,G(\mathbf{k},\omega),
  \qquad
  G(\mathbf{k},\omega)=\frac{1}{\omega-\varepsilon_{\mathbf{k}}
  -\Sigma(\mathbf{k},\omega)},
  \label{eq:green}
\end{equation}
where $\varepsilon_{\mathbf{k}}$ is the bare-band dispersion and $\Sigma=\Sigma'+i\Sigma''$ is the electron self-energy, whose real part $\Sigma'$ renormalizes the dispersion and whose imaginary part $\Sigma''$ sets the quasiparticle lifetime.
Extracting $\varepsilon_{\mathbf{k}}$ and $\Sigma(\mathbf{k},\omega)$ from the measured spectra, in addition to determining the overall shape of the band structure, is therefore a central inference goal of ARPES.

As recently emphasized by Borisenko, however, a fundamental yet often overlooked aspect of ARPES is that a measurement at fixed photon energy does not produce a simple energy momentum intensity map along a high-symmetry direction~\cite{borisenko2026}. Instead, photoelectrons with a given kinetic energy lie on an emission sphere. The familiar ARPES ``rainbow'' plot therefore represents a curved 2D slice through $(\mathbf{k},\omega)$ space rather than a true high-symmetry cut comparable directly to calculated band structures. Consequently, a conventional energy-distribution curve does not necessarily correspond to a single momentum.

In light of equation~\eqref{eq:intensity} and fundamental considerations above, opportunities for AI in ARPES workflow can be broadly grouped into three complementary roles: \textbf{\emph{data processing}}, such as denoising, artifact removal, and reconstruction of clean spectra; \textbf{\emph{inversion}}, aimed at extracting band dispersions, self-energies, or underlying Hamiltonian parameters from measured spectra; and \textbf{\emph{autonomy}}, in which AI guides data acquisition and instrument control. \textbf{\emph{Data processing}} and \textbf{\emph{inversion}} primarily support real-time and post-experiment analysis, whereas \textbf{\emph{autonomy}} operates predominantly during the sample fabrication and experiment to enable adaptive and ultimately self-driving measurements.

\section{Overview of ARPES workflow}
\label{sec:workflow}

\begin{figure}[htbp]
  \centering
  \includegraphics[width=0.9\textwidth]{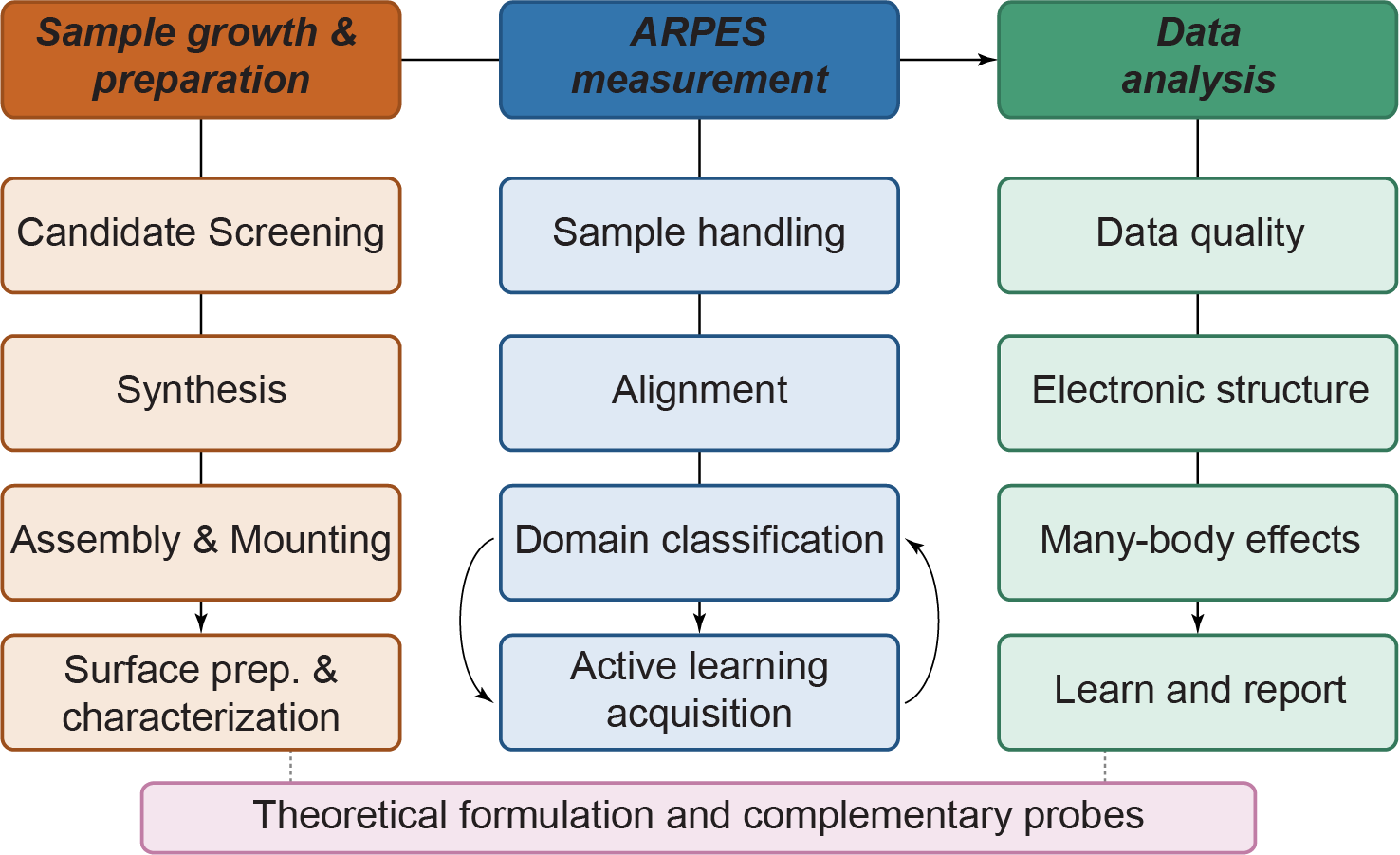}
  \caption{\textbf{Schematic of ARPES Workflow} The typical ARPES workflow progresses from left to right, from sample growth to analysis via measurement in an ultrahigh vacuum (UHV) chamber. Solid straight arrows indicate the nominal forward path, while solid curved arrows mark active learning–based iterations designed to acquire a rapid and complete dataset during a small experimental window. It represents an iterative closed loop between measuring, analyzing, and deciding the next step. Combined with other complementary studies, this reveals the underlying electronic properties of quantum materials}
  \label{fig:workflow}
\end{figure}

We place the ARPES experiment within a broader workflow, as illustrated in figure~\ref{fig:workflow}, which we divide into three major stages: \textbf{\emph {sample growth and preparation, \textup{ARPES} measurements, and data analysis.}} These stages are supported in parallel by theoretical modeling and complementary experimental probes, which guide material selection, experimental design, and interpretation of the measured electronic structure. ML frameworks and AI agents are increasingly being incorporated throughout this workflow to improve efficiency, accuracy, and degree of automation at each stage.

The \textbf{\emph{sample growth and preparation stage}} begins with identifying promising candidate materials and optimizing their synthesis. With the growing interest in 2D van der Waals (vdW) materials, this stage increasingly extends beyond bulk-crystal growth to exfoliation, assembly of heterostructures, and fabrication of functional microelectronic devices \cite{Mannix2022,masubuchi2018,Park2024}. The extreme surface sensitivity of ARPES requires ultra clean surfaces \cite{damascelli2003,cattelan2018perspective} which necessitates rigorous control over fabrication, contamination, and environment for preparing and studying air sensitive quantum materials. These considerations make sample preparation an important target for AI-assisted optimization and automation within the broader ARPES workflow.

The \textbf{\emph{measurement stage}} encompasses the entire ARPES experimental process, from sample handling and alignment to data and metadata acquisition and real-time evaluation of the measured spectra. Increasingly, these steps are interconnected through automated and active-learning approaches, in which preliminary measurements are used to identify or classify spatial domains, assess data quality, and determine where and how subsequent measurements should be performed \cite{agustsson2024,aims2026,na2025,melton2020}. Analysis therefore intersects directly with the measurement stage: each acquisition cycle may be followed by rapid processing, such as domain labeling, feature identification, or peak fitting to extract band dispersions, which can in turn inform the next acquisition cycle. This creates a closed-loop workflow in which measurement and analysis continuously inform one another \cite{pithan2023,agustsson2024,aims2026}.

Some of these \textbf{\emph {data analysis}} tasks naturally overlap with post-experiment data analysis, where more comprehensive treatment of the acquired datasets is performed. At this stage, the goal shifts from guiding data acquisition toward extracting the underlying electronic structure and ultimately understanding the physics governing the material property, including quasiparticle behavior and many-body interactions. The analysis stage also encompasses data curation, organization, additional metadata generation (associated with analysis and data reduction), and deposition into databases or open repositories, which are increasingly important for reproducibility and for enabling future data-driven approaches. AI and ML already appear at many nodes throughout this interconnected workflow \cite{chen2021review,pustovit2024,stach2021}. The following section therefore introduces the AI and ML terminology used throughout this review.

% , from sample preparation and autonomous acquisition to spectral analysis, physical interpretation, and data curation.

\section{AI terminology for experimentalists}
\label{sec:terminology}

The term AI, as commonly used today,  typically refers to artificial narrow intelligence (ANI)- systems designed to perform specific tasks, rather than the hypothetical artificial general intelligence (AGI). Throughout this review, we use AI as a broad, colloquial term for such task-specific computational intelligence. In experimental practice, it is useful to distinguish between conventional rule based algorithms and ML approaches. Conventional rule based symbolic AI relies on explicitly programmed instructions and typically requires substantial human intervention. This approach becomes increasingly challenging for large multidimensional 4D or 5D datasets acquired during a single experiment, combined with multimodal measurements for instance valence band and core level spectroscopy. The resulting complexity can overwhelm the conventional analysis pipeline and create a significant barrier between data acquisition and scientific interpretation. In contrast, ML methods are based on dynamic models that learn task-specific rules directly from the data by optimizing the model parameters, reducing the need for explicit programming of complex rules~\cite{jordan2015machine}. ML therefore provides a flexible framework for enabling \textit{data driven decision making} in an increasingly complex experimental landscape.

Traditional compact ML encompasses shallow models and conventional statistical models such as dimensional reduction using principal component analysis (PCA), non-negative matrix factorization (NMF) or $t$-distributed stochastic neighbor integration (tSNE), and clustering using \km\ or Density-Based Spatial Clustering of Applications with Noise (DBSCAN). In contrast, the deep learning model uses multiple nonlinear processing layers to automatically learn representations from complex multidimensional data. This approach has dramatically improved performance in tasks such as image and object recognition and shows considerable promise for analyzing spectroscopic maps~\cite{lecun2015deep}.

\begin{figure*}[htbp]
  \centering
  \includegraphics[width=0.9\textwidth]{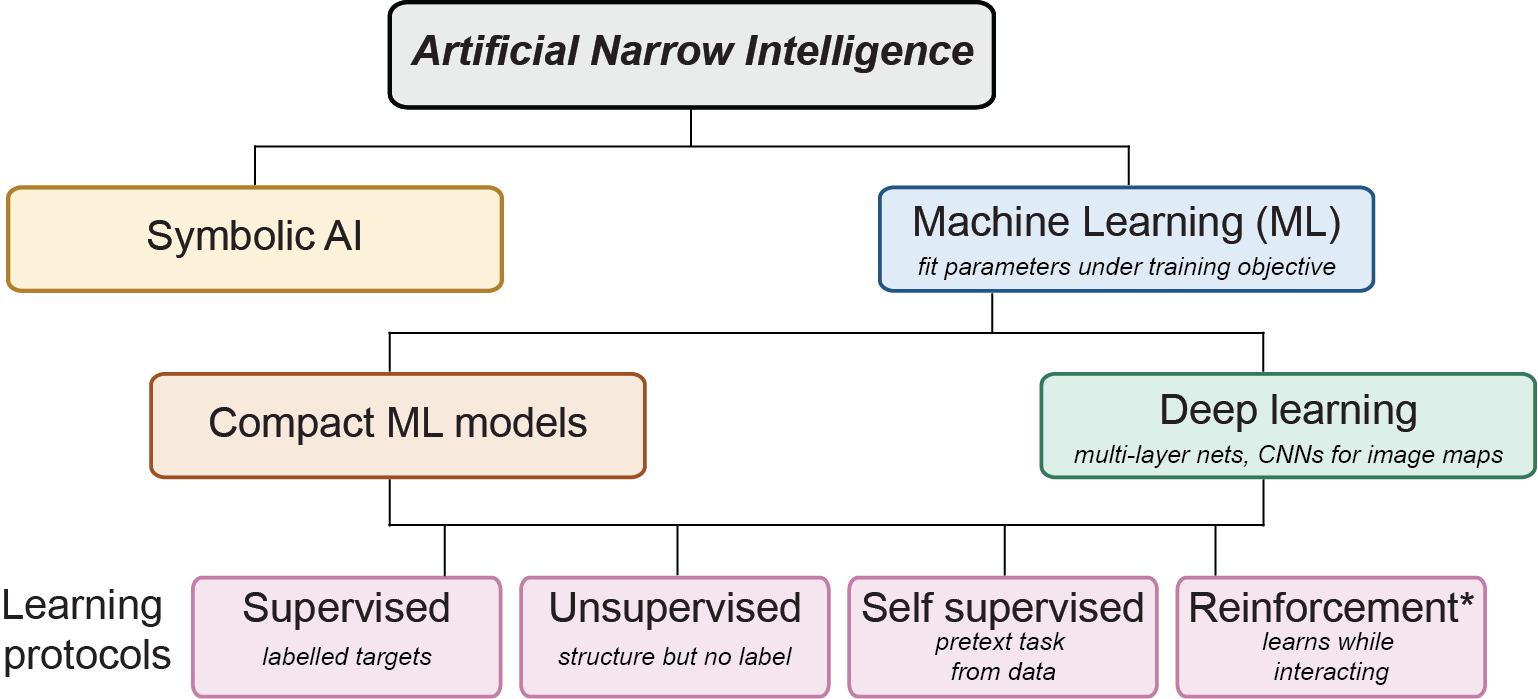}
 \caption{\textbf{A practical AI terminology for experimental physicists.} Machine learning splits by architecture into traditional compact models and deep learning networks. The four learning protocols represent a separate, independent axis of training models. The asterisk marks reinforcement learning, which in current practice is typically implemented with deep networks rather than as an architecture-independent protocol. The figure is a visual guide rather than a comprehensive glossary. Detailed examples appear in the text.}
  \label{fig:ai-term}
\end{figure*}

Independent of architectural design, data training protocols are commonly classified as supervised, unsupervised, self-supervised, or reinforcement learning \cite{jordan2015machine,lecun2015deep}. All of the statistical methods mentioned before are categorized under unsupervised learning method as it infers structure without external labels by grouping or compressing the raw data based on similarity. In contrast, supervised learning trains on labeled data as its ground truth. In ARPES, for example, dispersion along high symmetry direction might be marked ``good alignment" or a spectral feature labeled ``band". Self-supervised learning instead derives its own supervisory signals directly from unlabeled data through a user-defined pretext task. Examples include reconstruction-based approaches, such as denoising autoencoders that recover corrupted inputs~\cite{vincent2008}, and variational autoencoders used to reconstruct or compress ARPES maps~\cite{restrepo2022,autoencoder2025}, as well as contrastive and masked-image approaches, including DINO, SwAV, MoCo, BYOL, and related methods~\cite{caron2021dino,caron2020swav,he2020moco,grill2020byol}, as used in transfer-learning ARPES work~\cite{ekahana2023,graphconv2025}. Finally, reinforcement learning enables an agent to learn and maximize reward through interaction with an environment, updating the mapping from observations to actions without labeled targets~\cite{sutton1998reinforcement}. Such approaches are beginning to find applications in automated 2D-material fabrication~\cite{heteroassembly2026}. These commonly used AI terminologies are summarized in Figure~\ref{fig:ai-term}.

Beyond these learning paradigms, physics-informed neural networks (PINNs) incorporate governing physical equations directly into the loss function~\cite{raissi2019}, providing a framework for integrating physical constraints with data-driven learning and motivating differentiable forward models and physics-parameterized simulators~\cite{na2025,zhang2026}. Multimodal models integrate heterogeneous data types to extract complementary information, while generative models can create new data samples or representations \cite{wang2023scientific}. In scientific applications, however, predictions from generative models must ultimately be validated against experimental measurements, ensuring that physical conclusions remain grounded in measured rather than synthetic data. 

\section{Automation of sample preparation for ARPES measurements}
\label{sec:pre}

\begin{figure}[htbp]
  \centering
  \includegraphics[width=0.95\textwidth]{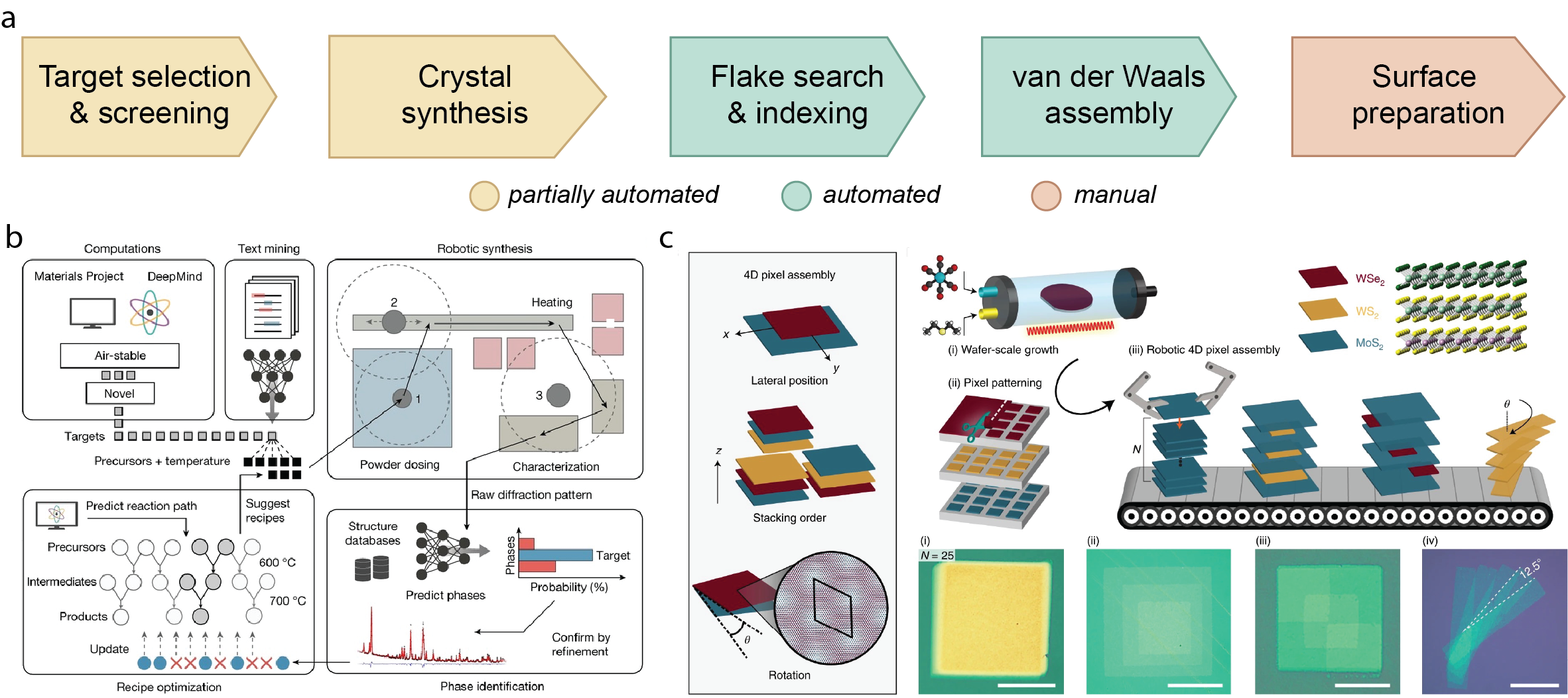}
  \caption{\textbf{Automated sample preparation and preliminary characterizations.} (\textbf{a})~The five pre experiment stages for ARPES experiment, shaded by the extent of automation. (\textbf{b})~An autonomous inorganic synthesis laboratory (A-lab) for air-stable targets, in which an ML algorithm proposes recipes, while robotic synthesis and learned phase identification form a self-revising loop (adapted from~\cite{szymanski2023}). (\textbf{c})~Robotic four dimensional pixel assembly, from wafer scale growth and
patterning to automated stacking, with control over layer number, composition,
lateral position and twist angle. Scale bar: 50 $\mu$m (adapted from~\cite{Mannix2022}).}
  \label{fig:pre}
\end{figure}

Before the ARPES measurement, several bottlenecks limit experimental throughput, such as bulk inorganic crystal synthesis, thin film deposition, surface preparation, 2D heterostructure fabrication, electrical contacting, sample mounting onto specialized platforms for applying electric or magnetic fields, current or strain. Figure~\ref{fig:pre}a summarizes this sample preparation workflow, color coded according to the current level of automation at each stage. 

A paradigm shift toward automated and ML-guided sample growth and preparation is underway \cite{stach2021,petkovic2025machine}. In bulk crystal growth, conventional trial-and-error screening is increasingly complemented by inverse materials design, in which deep learning models and high-throughput computational screening identify candidate materials that satisfy targeted property requirements~\cite{cheng2026materials}. A recent high-profile example is the A-Lab (Figure~\ref{fig:pre}b), which automates bulk solid-state synthesis through target selection, recipe generation, robotic synthesis, X-ray diffraction feedback, and iterative optimization of growth parameters~\cite{szymanski2023}. Operating on week-long timescales, the platform demonstrated the synthesis of dozens of inorganic materials in powder form, highlighting the potential of ML-guided planning to accelerate materials synthesis. Machine-learning approaches have also been applied more broadly to semiconductor and electronic-material growth, including both bulk-crystal and epitaxial-growth methods~\cite{petkovic2025machine}. For ARPES, however, extending such autonomous workflows beyond powder synthesis toward high-quality single crystals, epitaxial films, and well-ordered surfaces remains an important challenge.

For 2D materials and devices, the workflow presents a different set of challenges. Beyond selecting high-quality vdW source crystals, device fabrication requires exfoliation, identification of flakes of interest, deterministic pickup and stacking, cleanroom processing, electrical contacting, and integration onto ARPES-compatible sample holders. At the laboratory scale, substantial progress has been made toward automating individual steps of this workflow. Masubuchi~\emph{et~al.}~\cite{masubuchi2018} combined computer vision based flake detection with robotic pick-and-place assembly, enabling rapid identification and stacking with minimal human intervention. Mannix~\emph{et~al.}~\cite{Mannix2022} further advanced automated assembly through a robotic pixel-by-pixel approach for building vdW heterostructures under high-vacuum conditions (Figure~\ref{fig:pre}c). Roll-to-roll mechanical exfoliation concepts have also been proposed to reduce reliance on manual tape-peeling and increase flake throughput~\cite{Park2024}, while computer vision pipelines have improved flake detection, classification, and indexing for automated heterostructure assembly~\cite{flakeseg2025,heteroassembly2026}. Qumus extends these concepts to end-to-end automation using a miniaturized robotic platform  with multi-agent planning for flake isolation and vdW assembly, including graphene field-effect transistors (FETs) with closed-loop error correction~\cite{qumus2026}. Related platforms such as AIMS employ similar modular combinations of vision, robotics, and optional planning agents~\cite{aims2026}. Additionally, integrating active learning, semi-supervised learning, and ML models within a batch-wise architecture, Yan~\emph{et~al.}~\cite{yan2025large} introduced a self-evolving annotation ecosystem for flake segmentation.

Complementary infrastructure efforts, including the MonArk NSF Quantum Foundry~\cite{monark_foundry}, Argonne's RAPID-200 autonomous laboratory~\cite{anl_rapid200}, and the Henry Royce Institute UHV 2D materials platform~\cite{royce_uhv2d}, further support automated synthesis, assembly, and characterization relevant to ARPES sample preparation.

Together, these developments represent substantial progress toward automating the sample preparation side of the ARPES workflow. Nevertheless, surface preparation remains particularly critical and rate limiting step as ARPES requires atomically clean and flat surface. An important open question is therefore how effectively existing automated platforms can be generalized beyond commonly demonstrated graphene/hBN systems to the broader range of structurally complex condensed matter systems relevant to progress in quantum technologies, and how much expert intervention remains necessary throughout the complete preparation workflow.

\section{ARPES experiment and its automation}
\label{sec:experiment}

Once a sample is prepared, the experimental flow separates into two tasks. The first is \emph{sample transportation and positioning}: UHV transfer from the load lock to the measurement position, followed by alignment, orientation, focusing, and navigation across device-scale sample variations. The second is \emph{on-the-fly spectral measurement and next-point selection}: using spectra acquired across the spatial domain to decide where and what to measure next under UHV conditions.

Sample transport spans a spectrum of automation, from fully manual operation using wobble sticks, through motorized and computer-controlled motion, to closed-loop autonomous control on which software determines the next mechanical action. Motorization provides a practical bridge towards autonomy. The software controlled transfer arms, manipulators, stages, and cleaving mechanisms can be incorporated into AI driven workflows that monitor the experimental setup and determine subsequent actions. Motorization reduces manual intervention in routine sample transfer operations, cleaving, annealing, electrical testing and thermal metal evaporation. Expert effort then shifts from repetitive mechanical operations toward higher level decisions.

% , such as where and when a sample should be positioned and which experimental step should follow.

% For example, cleaving can be performed using a motorized mechanism rather than a manually operated wobble stick, while appropriately designed annealing chambers can minimize manipulation of transfer arm and valves.

The MAESTRO beamline at the Advanced Light Source provides a concrete example of motorized sample handling~\cite{maestrohighlight}. After a user inserts a sample into the load lock, transfer arms and sample trays can be moved under computer control from the loadlock towards cleaning/growth chambers and onto the measurement chambers. This motorized backbone represents an important step toward ARPES automation and provides the hardware infrastructure upon which future camera-based AI agents could monitor system states and guide experimental procedures. Facility roadmaps, including the AI@ALS workshop, similarly emphasize the need for software infrastructure, shared application programming interfaces (APIs), and streaming protocols that can connect motorized hardware with real-time data analysis and decision making~\cite{alsworkshop2024}.

\begin{figure}[htbp]
  \centering
  \includegraphics[width=0.9\textwidth]{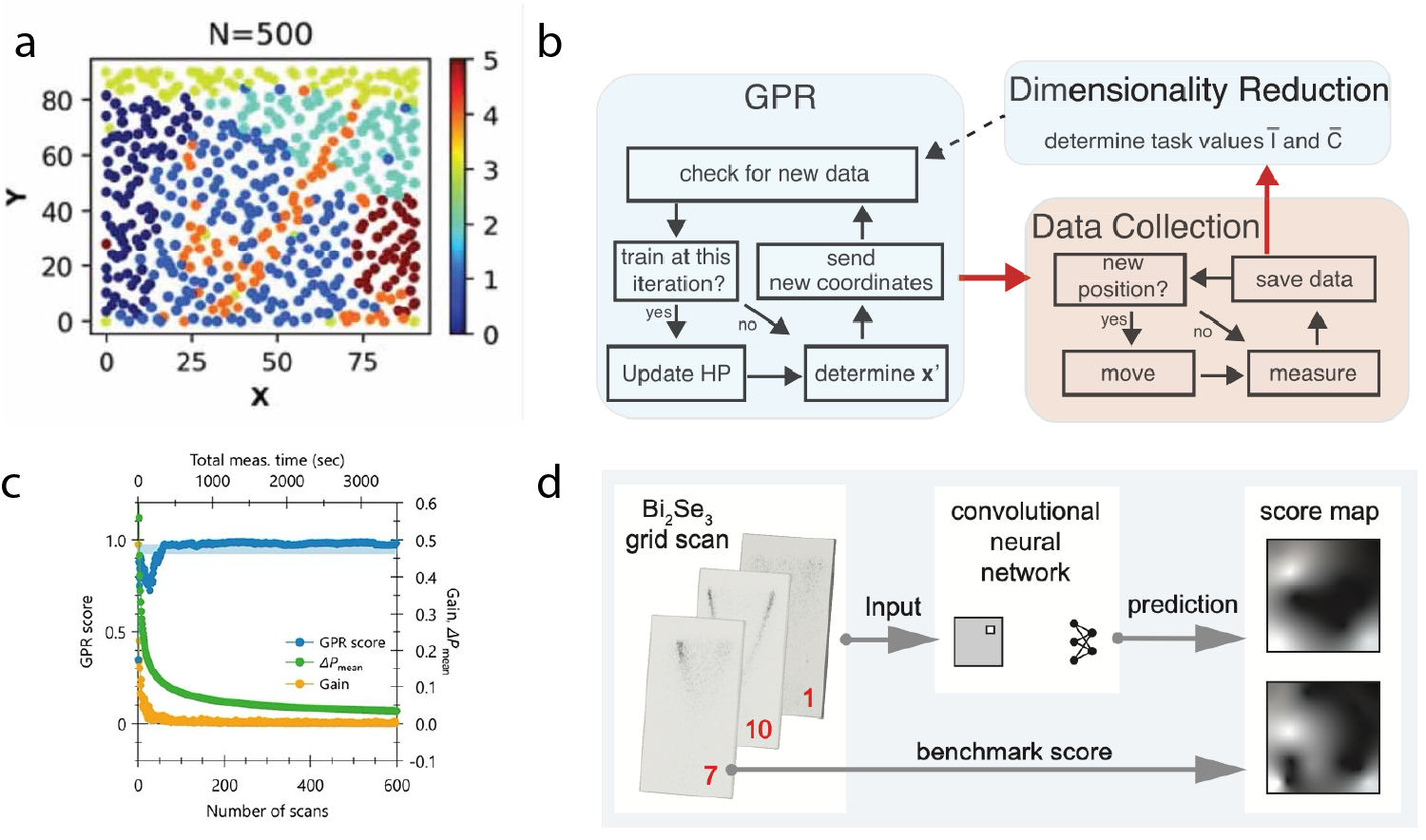}
  \caption{\textbf{Route to experimental automation}
  (\textbf{a})~KMGP-driven data collection identifying distinct major phases on the sample, colored by cluster label~\cite{melton2020}.
  (\textbf{b})~The GPR-guided acquisition loop showing the autonomous data acquisition process. The colored boxes represent asynchronously running processes, while the red arrows indicate real-time, ML-based communication between the instrument and Python scripts~\cite{agustsson2024}.
  (\textbf{c})~GPR score, mean uncertainty ($\Delta P_{\mathrm{mean}}$), and gain versus number of scans for Bi$_2$Te$_3$ spin-ARPES, illustrating a quantitative stopping criterion for the acquisition loop~\cite{iwasawaspin2024}.
  (\textbf{d})~A synthetic trained CNN scores the measured spectral quality across a Bi$_2$Se$_3$ grid scan. The predicted map is compared with a conventional benchmark score and decides where to measure~\cite{na2025}. Combined, these steps form a closed-loop experimental workflow: locating sample variations, selecting target regions, measuring until an objective stopping criterion is reached, evaluating spectral quality, and iterating to optimize spatial and electronic coverage.}
  \label{fig:experiment}
\end{figure}

With the sample positioned for measurement,  AI/ML can next address where and what to measure. Melton \etal~\cite{melton2020} developed KMGP by combining \kmc \ of the measured spectra with the Gaussian process regression on the resulting cluster labels to identify and map distinct electronic domains in the 4D parameter space (Figure~\ref{fig:experiment}a). However, this efficiency critically depends on cluster labels that faithfully track real spatial domains; when spectra vary continuously, or labels are inaccurate, sparse sampling may fail to recover the underlying physics. Related approaches place adaptive decision-making directly within the acquisition loop. \'{A}g\'{u}stsson, Hofmann and colleagues~\cite{agustsson2024} implemented a related guided data acquisition strategy at the SGM4 microARPES beamline (ASTRID2). Rather than following a predefined XY grid, the algorithm adaptively searches real and momentum space for regions containing high spectral intensity and sharp features. Figure~\ref{fig:experiment}b illustrates this closed-loop real-time approach, in which the ML-based analysis communicates with the beamline control software to determine subsequent measurements.
 
Adaptive acquisition can also improve the efficiency of particularly time-intensive measurements. Iwasawa, Ueno, and collaborators~\cite{iwasawaspin2024} applied GPR to spin-resolved ARPES using spin-polarization signals extracted from energy distribution curves (EDCs). They show that a GPR model enhances the spin ARPES measurement, significantly reducing the acquisition time cost by  5-10 times, compared to empirical stopping criteria. GPR score derived from EDCs provides a quantitative stopping criterion for the measurement loop. Figure~\ref{fig:experiment}c shows the GPR score, gain and mean uncertainty associated with spin polarization extracted from the spin-resolved EDCs in Bi$_2$Te$_3$. This illustrates how ML can optimize not only where to measure but also when sufficient information has been acquired. AI can further assist with sample alignment, mapping, and spectral-quality assessment.

Na~\emph{et~al.}~\cite{na2025} developed \textsc{aurelia}, an open-source, physics-parameterized synthetic-spectrum simulator capable of generating large training datasets. Coupled with a dual-branch convolutional neural network (CNN) trained exclusively on synthetic spectra, the framework enables rapid assessment of ARPES spectral quality without manually labeled experimental spectra, instead using simulator-generated training targets. The trained network can identify flake domains that show angular misalignment and locate spatially uniform regions suitable for measurement. Extended to rotational domains, the ML classifier can also determine relative orientations in twisted heterostructures. Such approaches can transform sample alignment and quality evaluation from slow, manually intensive procedures to higher-throughput workflows (Figure~\ref{fig:experiment}d). Important limitations nevertheless remain: matrix-element effects are randomized rather than explicitly calculated, spectral-quality labels rely on rule-based proxies, and the demonstration is restricted to a limited material system, requiring further validation across diverse ARPES datasets.

Once spatially resolved spectra are acquired, another challenge is to determine which measurements originate from the same electronic domain. Self-supervised image encoders pretrained in ImageNet, combined with few shot $k$-nearest-neighbor referencing, can automate this classification task~\cite{ekahana2023}. However, performance may be low when generic embeddings are clustered directly using \km\ without carefully selected reference examples. Graph convolutional networks operating on neighboring XY positions provide an alternative by classifying spatial information~\cite{graphconv2025}. By encouraging neighboring spectra to share band dispersion, graph based regularization improves spatial consistency while retaining spectral information, complementing expert verification of domain boundaries.

% rather than replacing

These examples demonstrate a progression from motorized hardware to adaptive closed-loop ARPES acquisition. However, their success depends critically on whether learned representations, cluster assignments, or optimization targets faithfully capture the underlying physics. When spectral properties vary continuously, cluster boundaries are ambiguous, or the chosen objective does not represent the scientifically relevant information, sparse or adaptive sampling can miss important electronic features. Reinforcement-learning strategies could extend these approaches by allowing an agent to update its measurement policy based on information gained from each acquisition. Ultimately, integrating motorized control, real-time analysis, uncertainty estimation, and adaptive decision-making while retaining expert supervision \& physically meaningful constraints forms the basis of robust autonomous ARPES within a specific scientific objective and limited beamtime. Recent workflows have successfully implemented multi-modal data fusion to co-analyze valence band and core-level spectra across wide spatial coordinates, leveraging Uniform Manifold Approximation and Projection (UMAP) to map continuous electronic and chemical phase landscapes with high visual fidelity~\cite{staab2024,staab2026}.

\section{Post-experiment data analysis and insights}
\label{sec:post}

ARPES experiments generate multidimensional datasets that contain rich information on electronic structure, such as discussed in Section~\ref{sec:arpes-basics}. Extracting this information requires systematic analysis that can be broadly divided into two complementary approaches: (1)  image-based analysis, which adapts methods from computer vision for denoising, feature identification, spectral enhancement, and spatial domain mapping~\cite{pustovit2024}, and (2) quantitative extraction of physical parameters, such as electronic dispersions, spectral linewidths,  self-energies, and energy gaps.

Two practical challenges affect both approaches. First, ARPES spectra can exhibit low signal-to-noise ratios (SNR), particularly under rapid acquisition conditions. Second, multi-band systems present substantial spectral-overlap ambiguity, where weak bands may be obscured by stronger, overlapping neighboring features~\cite{pustovit2024}. Consequently, denoising and feature separation, together with domain identification and classification, are important components of both real-time and post-experiment ARPES analysis.

\subsection{Denoising and spectral restoration}
\label{sec:denoising}

\begin{figure}[htbp]
  \centering
  \includegraphics[width=0.9\textwidth]{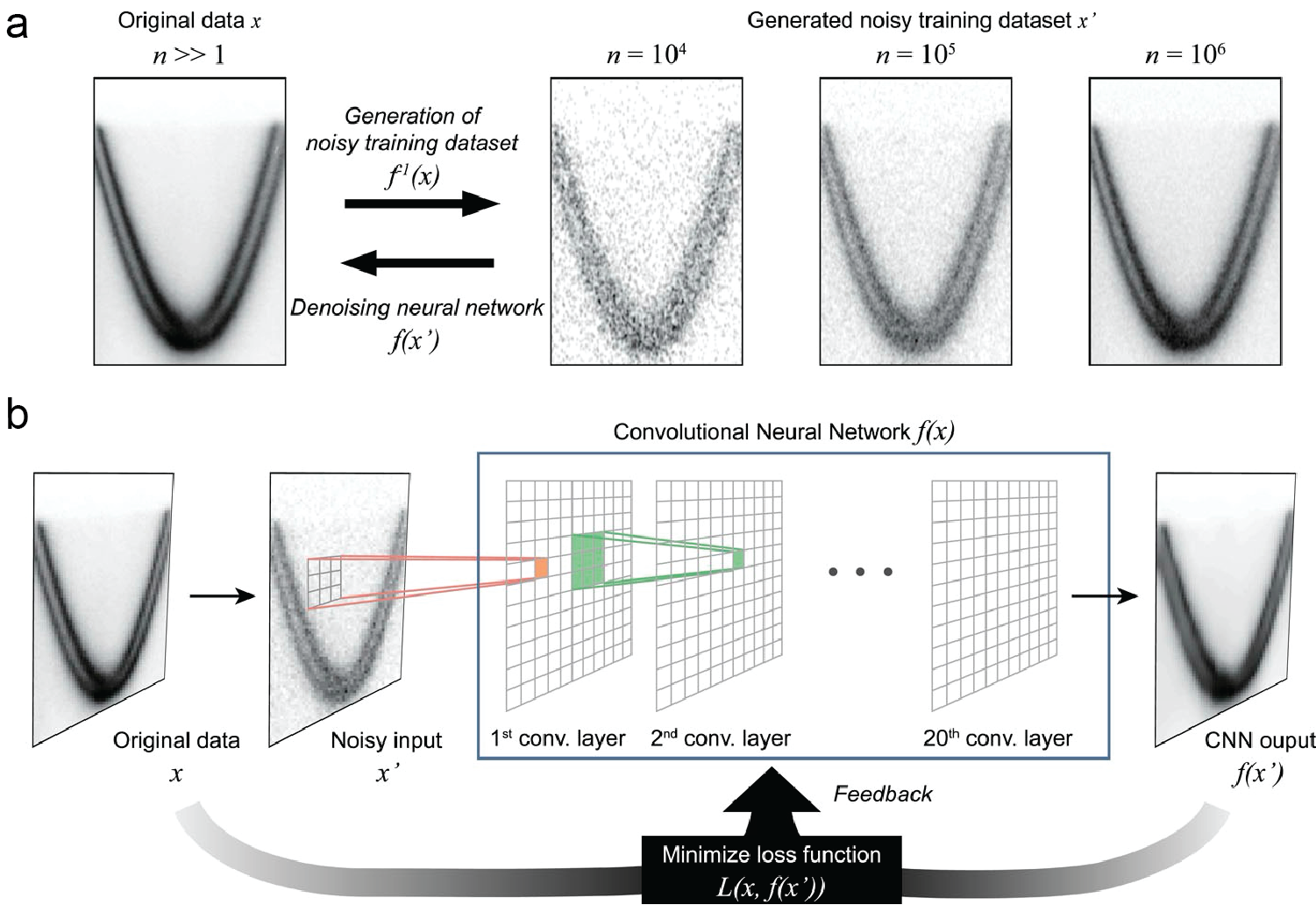}
 \caption{\textbf{Denoising network training sequence on Poisson subsampled ARPES data}
(\textbf{a})~Generation of noisy training data from high count Au(111) surface states spectra. From original high count data \(x\), low count noisy copies \(x'\) are generated via Poisson subsampling, simulating rapid acquisition. The denoising process is denoted \(f(x')\), while the generation of the noisy training set is denoted \(f^{-1}(x)\). 
(\textbf{b})~Training loop for the denoising neural network, where the loss function drives parameter optimization. The loss function $L(x,f(x'))$ combines mean absolute error (MAE) and multi-scale structural similarity index measure (MS-SSIM). The weighted formulation ensures that reconstructed spectra retain sharp spectral features essential for lineshape and second-derivative analysis, enabling quantitative comparison between short and long acquisitions on the same ARPES cuts (adapted from~\cite{kim2021}).
}
  \label{fig:denoise_kim}
\end{figure}

Machine learning methods offer a promising route to improve the quality of ARPES data while reducing the acquisition time. Kim~\emph{et~al.}~\cite{kim2021} validated ML-based denoising by comparing reconstructed short acquisition spectra with independently acquired long-exposure measurements of the same cuts. Importantly, performance was evaluated through the preservation of  quantities such as peak positions and linewidths rather than visual smoothness alone. Their network enabled second-derivative and line-shape analysis of spectra acquired with two orders of magnitude shorter dwell time. The overall training sequence is illustrated in Figure~\ref{fig:denoise_kim}, where low-count spectra are generated from high-count experimental data through Poisson subsampling to reproduce the counting statistics of rapid ARPES measurements, and a CNN is trained to reconstruct the corresponding high-count spectra.

Restrepo~\emph{et~al.}~\cite{restrepo2022} instead trained a shallow convolutional variational autoencoder (VAE) using  synthetic spectral function maps incorporating Gaussian resolution broadening and Poisson noise. Applied to experimental ARPES data, a network trained for Bi2212 suppresses modeled noise in low SNR dispersion maps, while a separate model for 1T-TiSe$_2$ performs both denoising and peak sharpening following artificial broadening. Because the training targets are generated synthetically rather than obtained from long-dwell experimental times, performance depends critically on how  accurately the simulations reproduce experimental noise, resolution, and spectral line shapes. This consideration becomes particularly important when the network performs sharpening in addition to noise removal.

Liu~\emph{et~al.}~\cite{liu2023} addressed a different experimental artifact: the grid-like pattern introduced by a physical collimator positioned in front of the charge-coupled device (CCD) detector to block stray photoelectrons. Conventional Fourier filtering cannot completely remove this structure because the artifact is not simply linearly superimposed on the intrinsic spectrum, and aggressive filtering can remove genuine spectral information. Using self-correlation information within the spectra to parameterize the measured data as a superposition of three nearly independent components (the intrinsic energy bands, Poisson noise, and the grid artifact), three coupled neural networks jointly disentangle these components and simultaneously suppress both grid and noise contributions while preserving the underlying band structure and spectral sharpness. This approach allows rapidly acquired ARPES spectra to approach the scientific utility of substantially longer swept-mode measurements.

The high cost of acquiring large experimental training datasets remains an important obstacle to deep-learning approaches. Cheng~\emph{et~al.}~\cite{prb2026} addressed this limitation through a Sim2Real strategy based on a gradient attention U-Net trained using physics-based simulations and data augmentation. The approach learns characteristic ARPES noise without requiring extensive training in experimental data and was shown to preserve quantitative band dispersions and spectral features, remove detector-induced grid artifacts while retaining low-frequency spectral information, and achieves up to an approximately 500-fold reduction in acquisition time for useful spectra.

Yokoyama~\emph{et~al.}~\cite{deepprior2026} pursued a complementary, training-free strategy based on the deep image prior, which leverages the inherent inductive bias of convolutional architectures to reconstruct spectral structure without a pretrained dataset. This method achieves an approximately 70-fold reduction in acquisition time for both quantitative EDC/MDC (momentum-distribution curve) analysis and up to a 300-fold reduction for qualitative band structure determination. Such training-free methods are attractive for ARPES because they avoid the need to construct large labeled datasets and can potentially be deployed across experiments with relatively modest computational overhead.

Denoising approaches are particularly valuable in several aspects as follows. The soft and hard X-ray ARPES regimes can benefit from it, where poor energy resolution or low photon flux limits reliable band-parameter extraction. In particular, this method is beneficial for time-resolved experiments, and fragile strongly correlated systems (2D quantum devices), where complex self-energy dependent line shapes would otherwise require prohibitively long integration times to resolve. Once trained and deployed, deep learning models are computationally efficient during inference, requiring minimal GPU resources and enabling real-time processing. The hardware independent analysis is a scalable next step for high-throughput data processing across diverse experimental conditions and will evolve over time as more data is generated.

However, the reported acquisition-time reductions should be interpreted within the validation conditions of each study rather than as universal replacements for high-statistics measurements. The central criteria are not whether a processed spectrum appears smoother, but whether physically extracted quantities remain quantitatively reliable and whether decisions on what to scan next may be reliably made. This distinction is important because both manual image processing and automated ML approaches have inherent limitations. Manual processing can introduce  subjective choices and expert bias, while ML-based restoration can over-smooth spectra details and complicate accurate band identification. Meanwhile, denoising may introduce false features when bands are intrinsically broad or poorly characterized~\cite{restrepo2022,kim2021,prb2026}. In this case, a validation against longer reference measurements or independent analysis methods is therefore essential before scientific conclusions are drawn from processed spectra~\cite{kim2021,prb2026}. Algorithmically enhanced or reconstructed spectra should also be explicitly identified as processed data, with the quantitative validation procedure clearly documented.

\subsection{Spatial-domain identification and clustering}
\begin{figure}[htbp]
  \centering
  \includegraphics[width=0.95\textwidth]{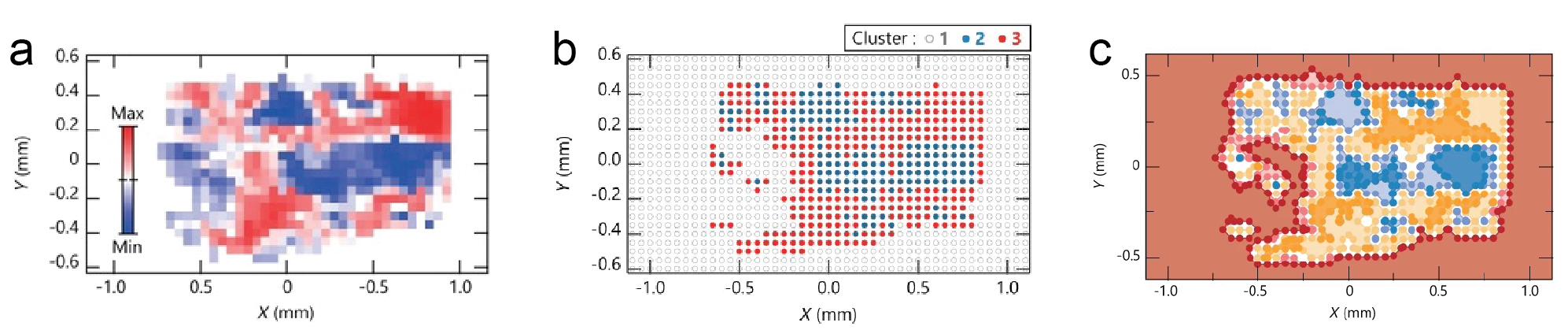}
 \caption{\textbf{Unsupervised clustering for spatial domain identification} Panel (a) shows raw spatially resolved spectral intensity. Panel (b) applies \kmc\ with three clusters (1,2,3), assigning each point exclusively to one cluster. Panel (c) applies fuzzy c-means soft clustering, where each point receives probabilistic membership weights across clusters; the background shading indicates low membership confidence or non-sample regions, revealing sharp domain boundaries (adapted from~\cite{iwasawa2022})}
 \label{fig:cluster}
\end{figure}

Unsupervised clustering provides a powerful approach for converting spatially resolved ARPES datasets into electronic-domain maps~\cite{melton2020,staab2024,staab2026}. Iwasawa~\emph{et~al.}~\cite{iwasawa2022} apply \km~\cite{macqueen1967multivariate} and fuzzy $c$-means~\cite{bezdek2013pattern} clustering to spatially resolved ARPES measurements of YBa\(_2\)Cu\(_3\)O\(_{7-\delta}\), automatically distinguishing  surface terminations through the differences in their electronic structures. Figure~\ref{fig:cluster} compares the resulting cluster maps with conventional window-integration analysis.

Single pass clustering may nevertheless miss subtle spectral variations,particularly in spatially heterogeneous vdW heterostructures. The Multi-Stage Clustering Algorithm (MSCA)~\cite{msca2024} addresses this limitation by sequentially applying \(k\)-means first to selected energy--momentum windows in real space, and subsequently to the resulting spatial metrics in momentum space. This procedure identifies spectral regions exhibiting the strongest electronic inhomogeneity. Non-negative matrix factorization (NMF) provides an alternative path, by decomposing the full spectral dataset into physically interpretable basis vectors (e.g., distinct band dispersions or phase contributions) and activation weights~\cite{imamura2024}. Subsequent \kmc\ of the activation vectors can reveal the spatial distribution of different electronic components and can effectively separate mixed-phase spectral contributions within a single beam spot, beyond a simple pixel-wise classification.

Determining optimal clustering parameters remains challenging. \km ++, agglomerative clustering, DBSCAN, and related approaches may provide advantages for particular datasets, but no clustering algorithm removes the need for physical interpretation. Automatically identified clusters must ultimately be associated with physically meaningful domains rather than treated as ground truth. This becomes particularly important when spectral contrasts are weak, domains are characterized by strong angular spread in their orientations, or domain boundaries are continuous, because different algorithms or hyperparameters can produce different spatial partitions from the same dataset.

\subsection{Band Extraction and low-dimensional representations}
\begin{figure}[htbp]
  \centering
  \includegraphics[width=0.9\textwidth]{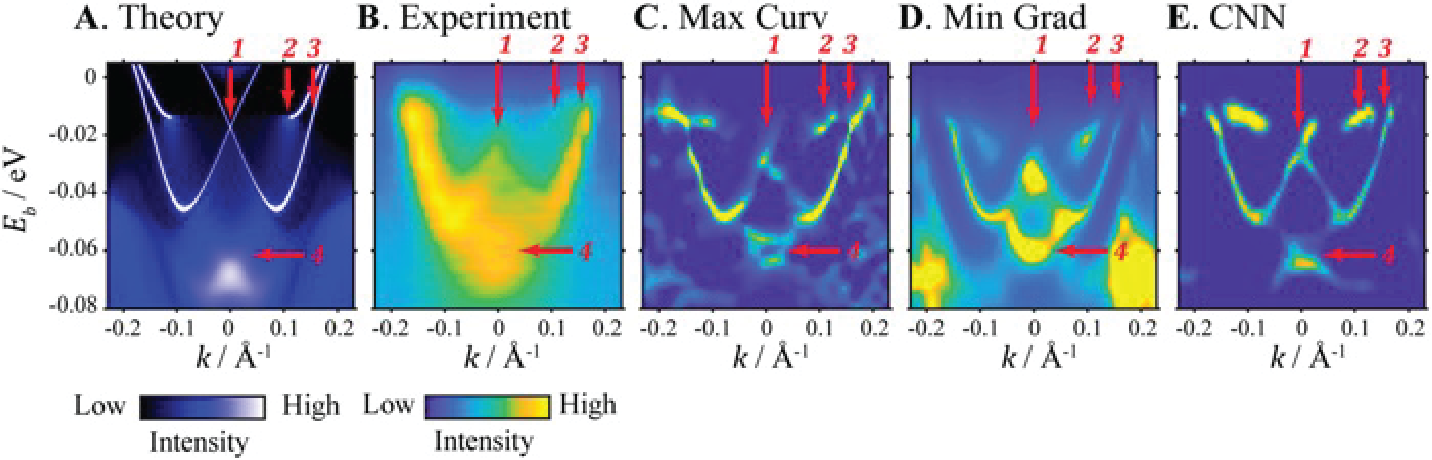}
  \caption{\textbf{Band extraction of NbIrTe$_4$ using DFT, conventional analysis, and machine learning.} (\textbf{a})~Density functional theory calculation of the band structure of NbIrTe\(_{4}\). The bands crossings are marked with red arrows. (\textbf{b})~Corresponding experimental ARPES data showing the same band structure. (\textbf{c--e})~Results from three band extraction methods applied to the experimental data: maximum curvature tracing (\textbf{c}), minimum gradient method (\textbf{d}), and CNN (\textbf{e}). The CNN approach recovers band positions comparable to conventional methods while reducing manual intervention and improving robustness on weak or overlapping features (adapted from Ref.~\cite{peng2020}).}
  \label{cnnpeng}
\end{figure}

Converting ARPES intensity maps into explicit quasi-particle dispersions remains challenging when analysis relies only on local intensity maxima. Peng~\emph{et~al.}~\cite{peng2020} applied a super-resolution CNN (SRCNN) to spectroscopic maps to enhance spectral contrast and improve band extraction relative to conventional interpolation and derivative-based approaches (Figure~\ref{cnnpeng}). More broadly, this work illustrates how feature extraction from ARPES maps can be formulated as an image-restoration problem, allowing mature computer-vision techniques and CNN architectures to be adapted to electronic spectroscopy.

Although ARPES datasets can occupy a high-dimensional measurement space, many physically relevant quantities, such as band positions, linewidths, energy gaps and self-energies, typically occupy a substantially lower-dimensional representation. The ARPESNet, developed by Ágústsson~\emph{et~al.}~\cite{autoencoder2025}, uses an autoencoder to compress ARPES spectra into a compact latent representation while retaining sufficient information for high-quality reconstruction. Such unsupervised representations provide an efficient means of data compression and feature extraction without explicitly specifying the underlying physical parameters. Complementary to latent-space compression, Xian~\emph{et~al.}~\cite{xian2023} formulated band extraction as a probabilistic band reconstruction problem rather than pointwise intensity tracing. Band loci are represented as a connected field using Markov random field / Bayesian maximum-a-posteriori (MAP) estimation, initialized from semi-local DFT calculations and constrained by a nearest-neighbor Gaussian prior enforcing band continuity. Applied to WSe\(_2\), the method reconstructs all 14 valence bands and recovers momentum-dependent structure that is difficult to obtain through independent pointwise fitting. The theoretical prior stabilizes the reconstruction in weak-signal regions but can also bias the solution toward the initial band structure, emphasizing the need to distinguish information inferred from the measurement from information imposed by the prior. Unsupervised approaches also extend naturally beyond equilibrium measurements. Majchrzak~\emph{et~al.}~\cite{majchrzak2025} applied \kmc\ to time-resolved ARPES on PtBi\(_2\), to track carrier dynamics and identify differences in relaxation behavior between states near the Weyl points and the bulk Fermi surface. 

\subsection{From spectra to many-body parameters}
\begin{figure}[htbp]
  \centering
  \includegraphics[width=0.75\textwidth]{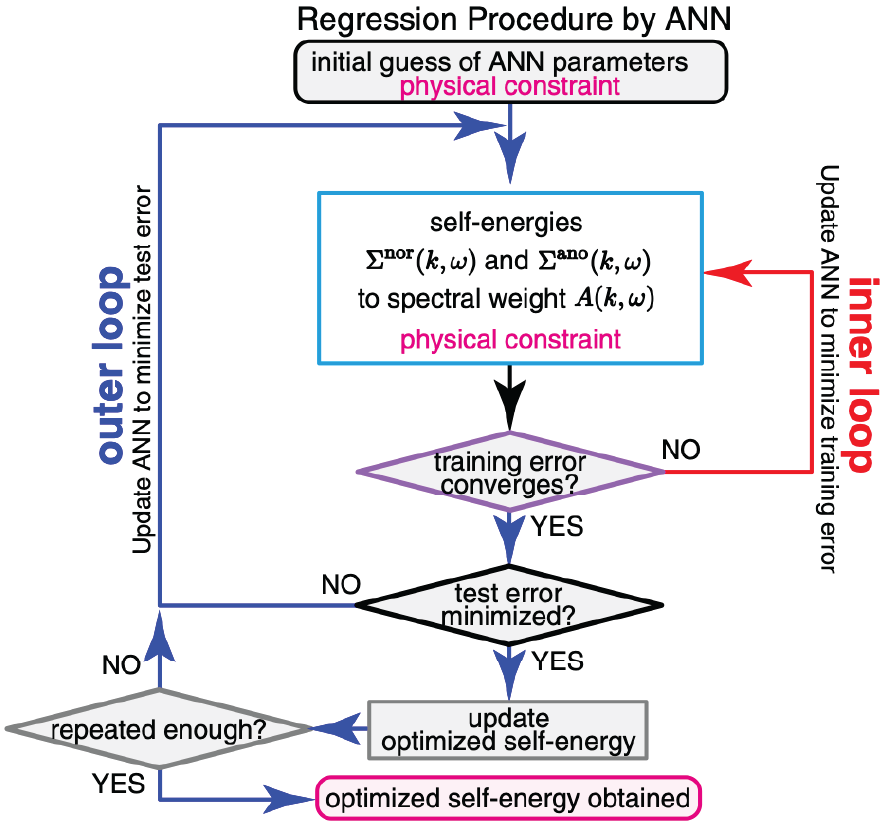}
  \caption{\textbf{Flow chart of an ML regression procedure for extracting hidden self energies in Bi2212.} A Boltzmann machine represents the normal (\(\Sigma^{\mathrm{nor}}(k,\omega)\)) and anomalous self energies ( \(\Sigma^{\mathrm{ano}}(k,\omega)\)), and is optimized against the measured spectral weight \(A(k,\omega)\) through nested training and test loops until convergence. Adapted from Ref.~\cite{yamaji2021,yamaji2023erratum}}
  \label{fig:yamajiml}
\end{figure}

A more demanding post-experiment objective is to infer physical quantities that are not directly visible in the measured intensity. Yamaji \emph{et~al.}~\cite{yamaji2021,yamaji2023erratum} employed Boltzmann machine regression to extract normal and anomalous self-energy components from ARPES spectra of the cuprate superconductors Bi2201 and Bi2212.  The ML algorithm revealed pronounced energy-dependent peaks in both self-energy components that largely cancel in the total spectral function, making them difficult to identify through conventional analysis. Remarkably, the anomalous self-energy peak was found to account for more than 90\% of the superconducting gap, identifying it as a dominant contribution to high-$T_c$ superconductivity. The method was benchmarked against simple metals, conventional BCS superconductors, and strongly correlated model systems with known solutions. The analysis further connects the superconducting transition temperature to the interplay among superfluid density, effective interaction strength, and Planckian dissipation. This work illustrates how ML-based inversion can uncover many-body information hidden in experimental spectra while relaxing assumptions commonly used in conventional analysis, such as linear bare dispersions and momentum-independent self-energies. Nevertheless, reliable inversion requires high-SNR ARPES data and benefits from independent physical validation.  Figure~\ref{fig:yamajiml} summarizes the ML-regression workflow and the extracted hidden self-energy components.

Complementary Bayesian maximum entropy tools implemented in \textsc{xARPES}~\cite{xarpes2026} extend self-energy and Eliashberg function extraction to systems with curved dispersions, jointly optimizing parameters describing both interacting and non-interacting dispersions. The authors show that traditional Lorentzian fitting of MDCs, suitable for linear dispersions, introduces systematic bias when applied to curved dispersions, precisely the regime relevant to analyzing dispersion kinks from electron-phonon interactions. The freely available, GPLv3-licensed Python code \textsc{xARPES} complements existing analysis frameworks and operates in two stages: first, the band map is quantified from raw intensity data via non-interacting dispersions and causal self energies, with the user specifying the number of dispersions and self-energy parameterizations, optionally complemented by a photoemission matrix element; and second, the total self-energy is decomposed into impurity, electron-electron, and electron-phonon contributions, while simultaneously extracting the electron-phonon Eliashberg coupling function. This provides a platform for accurately comparing experimentally extracted spectral functions against first-principles calculations in the presence of many-body interactions, particularly electron-phonon coupling. Demonstrations include phonon mode identification in a 2D electron liquid on TiO\(_2\) terminated SrTiO\(_3\) and consistent Eliashberg functions extraction~\cite{xarpes2026}.

The scientific goal may extend beyond dispersions to the Hamiltonian parameters that generate them. Building on implicit neural representation (INR) approaches introduced for inelastic neutron scattering~\cite{chitturi2023}, Zhang~\emph{et~al.}~\cite{zhang2026} use a sinusoidal representation network (SIREN) as a differentiable surrogate for the electronic structure of perovskite nickelates and manganites. Automatic differentiation is then used to optimize the tight-binding parameters against measured ARPES intensity, providing a route toward extracting underlying Hamiltonian parameters while bypassing conventional peak fitting. The quantitative agreement between the SIREN surrogate and measured ARPES intensity demonstrates that this ML-driven approach can automate data reconstruction and offer a promising connection between multi dimensional experimental and microscopic physical parameters.

\subsection{Transfer learning and future autonomous analysis}
The above approaches require extensive labeled training data that remain scarce and are not centrally available, motivating alternative strategies that reduce dependence on labeled experimental datasets. Transferable neural networks trained primarily on simulated spectra and transferred to scarce labeled experiment, can accurately classify the superconducting state of Bi2212 from a single snapshot~\cite{chen2025dann}. The result suggests a powerful new tool for identifying thermodynamic phase transitions and long-range order from electronic spectroscopy, even when the discriminating features do not conform to traditional gap based analysis.

Taken together, these methods form a modular ARPES analysis toolkit rather than a single universal ML solution. A future direction is to replace iterative optimization  with a neural network that directly infer model parameters, using standardized analysis pipelines to generate harmonized training datasets~\cite{xarpes2026}. More broadly, spectroscopy-specific models could be integrated with general purpose reasoning tools, enabling AI systems to autonomously identify relevant degrees of freedom, lattice symmetries, matrix-element effects rather than requiring expert input. Such integration is already emerging through agentic developments like \texttt{erlabpy}~\cite{han_erlab} and \texttt{ARPESskill}~\cite{ekahana_arpesskill}, which equip AI agents with the programmatic tools and domain skills needed to reason about analysis strategies rather than relying on rigid, predefined models.

\section{Theory support: DFT surrogates and ARPES}
\label{sec:theory}

Interpretation of ARPES measurements relies heavily on comparison with theoretical band structures. However, conventional self-consistent-field (SCF) calculations are generally computationally too expensive to provide real-time feedback during the ARPES beamtime and perhaps also too slow for an immediate post-experiment analysis. The role of AI is therefore not to replace the underlying physics of Density Functional Theory (DFT), but to shift much of the computational cost to the training stage, allowing learned surrogate models to generate electronic structures at a pace more compatible with ARPES experiments. Importantly, speed does not guarantee the physical accuracy of the quasiparticle bands. Because these models learn from existing electronic-structure calculations, their predictions inherit the approximations and biases of the underlying training data. Broader roadmaps detailing specific architectures, datasets, and common errors associated with the balance between speed and accuracy are described in detail in~\cite{kulik2022}. From this need, multiple groups developed similar AI models independently; the overview below highlights the representative approaches rather than claiming strict historical priority.

Early approaches demonstrated that ML could bypass parts of computationally demanding electronic-structure calculations by directly predicting charge densities, densities of states (DOS), or electronic operators~\cite{chandrasekaran2019, schnorb2019}. Chandrasekaran, Ramprasad, and colleagues successfully mapped local atomic environments onto charge density and the local DOS, allowing reconstruction of the DOS without requiring a Kohn–Sham solve during inference~\cite{chandrasekaran2019}. In parallel, Schütt \emph{et al.} introduced SchNOrb to predict molecular Hamiltonians and orbitals, demonstrating that electronic spectra could be derived directly from a learned operator~\cite{schnorb2019}. Although this initial demonstration was restricted to molecules rather than the crystalline solids, this operator-based route laid crucial groundwork for Hamiltonian learning approaches for crystals. Ben Mahmoud \emph{et al.} subsequently modeled the condensed-matter DOS as a sum of atom-centered contributions trained on DFT eigenvalues, enabling rapid estimations for large systems where direct SCF calculations are prohibitively expensive~\cite{benmahmoud2020dos}. More recently, How \emph{et al.} generalized this DOS prediction approach across diverse chemistries using the PET-MAD-DOS framework, extracting the Fermi level and bandgap as side products while allowing for optional fine-tuning~\cite{how2025petmaddos}. However, while these DOS and density models rapidly provide vital electronic-structure context, they do not produce the momentum-resolved band maps~\cite{kulik2022} required for ARPES. Generating these necessary $E$--$k$ dispersion overlays requires dedicated Hamiltonian or quasiparticle surrogates, which are discussed below.

For crystalline solids, including  vdWs, Hamiltonian learning approaches provide a more direct route to momentum-resolved bands. The DeepH framework learns the DFT Hamiltonian using a message-passing neural network formulated in a localized-orbital basis~\cite{deeph2022}. By exploiting local coordinates, DeepH effectively manages dimensionality, rotational variance, and gauge issues while bypassing the computationally expensive SCF cycle during inference~\cite{deeph2022}. The model achieves meV-scale accuracy in its Hamiltonian predictions, which is promising for generating DFT-level $E$--$\mathbf{k}$ band maps for comparison with ARPES~\cite{deeph2022}. However, it fundamentally remains a DFT-level surrogate, meaning its physical accuracy is capped by the underlying DFT training data unless a higher-fidelity teacher model is employed~\cite{deeph2022}. Subsequent iterations of DeepH have further expanded this foundation, introducing strict equivariance (DeepH-E3) and universal Hamiltonians, alongside new support for magnetism, hybrid functionals, and plane-wave bases~\cite{deephe3_2023,deephuniversal2024,deephmagnetic2023,deephhybrid2024,deephplanewave2024}.

A complementary strategy targets the exceptionally large supercells encountered in twisted vdW heterostructures.  Lou, Lewis, and Rossi bypassed full superlattice ab initio calculations entirely by training long range density models on comparatively small bilayer cells and extrapolating them to much larger twisted moiré supercells~\cite{lou2026moire}. This approach makes otherwise expensive moiré calculations tractable and reproduces characteristic electronic features, including flat band formation, in systems such as graphene, hexagonal boron nitride (hBN), and transition metal dichalcogenides (TMDCs). However, the extrapolation from small training cells to large moiré structures introduces an additional approximation that must be carefully validated, particularly when comparing with the narrow and strongly twist-dependent bands resolved by ARPES.

\begin{figure}[htbp]
  \centering
  \includegraphics[width=0.9\textwidth]{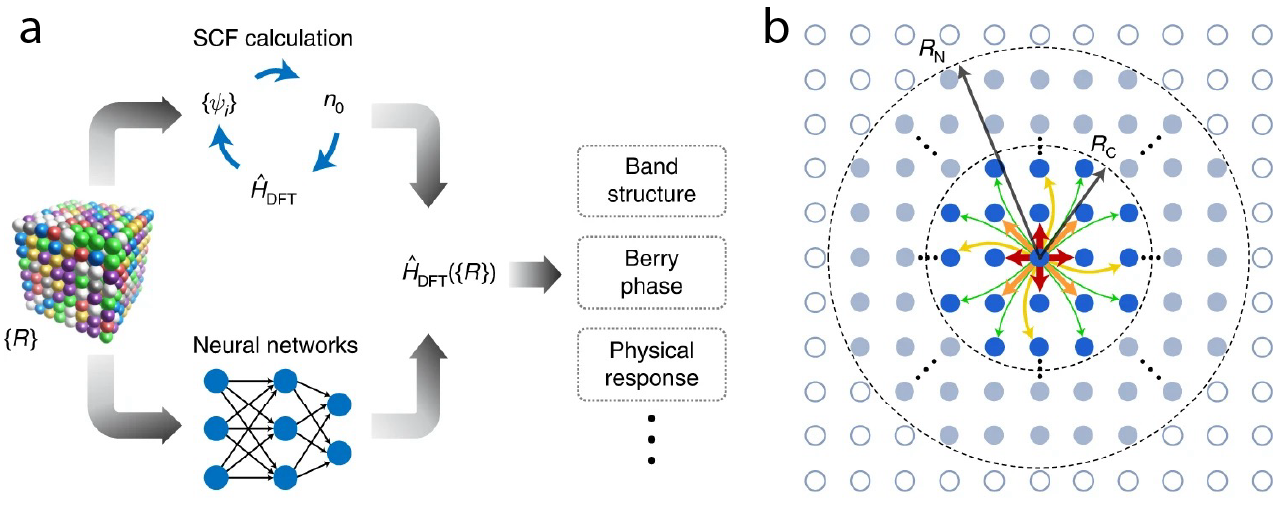}
  \caption{\textbf{DeepH: An example of a theoretical framework where a DFT Hamiltonian surrogate predicts the Hamiltonian from the atomic structure.} Adapted from Ref.~\cite{deeph2022}. (\textbf{a})~The DFT Hamiltonian $\hat H_{\mathrm{DFT}}(\{\mathcal{R}\})$, which is a function of atomic coordinates, can be obtained by self-consistent field (SCF) calculations or learned by a neural network for efficient \textit{ab initio} electronic structure calculations, yielding its physical properties (\textbf{b})~The network exploits the nearsightedness principle of electronic matter making the surrogate size transferable. The Hamiltonian matrix elements in the localized basis are nonzero only between neighboring atoms (within a cutoff $R_C$) and are influenced only by the local neighborhood (within $R_N$).}
  \label{fig:dft}
\end{figure}

An additional challenge is that ARPES probes interacting quasiparticle excitations, whereas standard DFT eigenvalues do not generally represent quasiparticle energies. ML can help bridge this gap by learning higher level many body corrections. Kn{\o}sgaard and Thygesen developed a method to fingerprint electronic states (ENDOME / RAD-PDOS) and train gradient boosting on tens of thousands of $G_0W_0$ energies to predict self-energy corrections at roughly $0.14$\,eV mean absolute error (MAE), giving GW-trained corrections at DFT evaluation cost~\cite{knosgaard2022}.  Hou~\emph{et~al.} instead used unsupervised VAE representations of Kohn-Sham states and predict GW quasiparticle structures for 2D metals and semiconductors, reporting errors of about $0.11$\,eV MAE~\cite{kohnshamvae2024}. These approaches illustrate how the computational cost of higher-level quasiparticle calculations can be partially transferred to a trained surrogate, although their accuracy and transferability remain determined by the quality and coverage of the reference calculations.

Constrained tight-binding models provide another complementary route. SlaKoNet learns Slater-Koster matrix elements from JARVIS-DFT data across a broad range of elements keeping tight-binding interpretability while improving gaps relative to GGA (generalized gradient approximation) and increasing evaluation speed on GPUs~\cite{slakonet2025}; a transparent DFT-level band model, not a GW (Green's function $\times$ screened Coulomb $W$) surrogate.

For electronic structure study, these surrogate models offer several practical opportunities during and after experiments. First, calculated dispersions could be generated rapidly and overlaid with measured spectra during or immediately after acquisition, providing direct comparison with bands reconstructed from the experimental data using the approaches discussed in Refs.~\cite{peng2020,xian2023,yamaji2021,xarpes2026}. Rapid calculations could also assist the interpretation of  $k_\perp$-dependent measurements, including identification of Brillouin-zone boundaries and estimation of the inner potential $V_0$. Second, theoretical predictions could, in principle, enter the acquisition loop itself. For example, a surrogate model could suggest the photon energy required to access a predicted feature such as a Weyl point, or help identify experimental conditions such as photon polarization that maximize sensitivity to particular states. Such real-time theory-guided acquisition remains less established than post-experiment applications but represents an important direction toward closed-loop ARPES.

Finally, surrogate models are particularly attractive when the structural hypothesis evolves during analysis. Electronic structures could be regenerated rapidly as external parameters and structural degrees of freedom are varied, without repeating a complete ab initio calculation for every candidate structure. Coupling such fast theoretical surrogates with real-time ARPES analysis could therefore transform theory from a primarily post-experiment interpretive tool into an active component of experimental decision making.

%% ====================================================================
\section{Data landscape: sporadic open repositories versus missing centralized ARPES database}
\label{sec:data}

The AI methods surveyed in the preceding sections share one common practical bottleneck during training: the ARPES community currently lacks a large, shared, and task-labeled ARPES dataset similar in scale and reusability to ImageNet. The idea of an “ARPES ImageNet” highlights the need for a common and well-documented repository that independent groups can use to train and benchmark their models. However, this does not mean that ARPES data can be treated simply as 2D images. ARPES datasets contain information about the experimental conditions, instrumental settings, underlying physics, and multiple experimental axes. Therefore, the labels required for machine learning strongly depend on the specific task. These tasks may include paired images for denoising or degridding, domain labels for spatial maps, target quantities for  band or self-energy extractions, and alignment scores for real-time data acquisition. For example, a denoising model trained on raw, Poisson-limited counts is solving a different problem from a model trained on second derivative images. These tasks also use data at very different stages of processing. It is therefore important to clearly separate raw detector events, calibrated intensities mapped into energy momentum coordinates, processed data that include normalization, background subtraction, interpolation, or derivatives, and finally, human- or model-generated labels.

At present, the open data landscape in ARPES remains fragmented. Public data are often deposited only when required by journals, funding agencies, or institutions, while most research groups maintain much larger private archives on institutional storage. A complete survey of all publicly available ARPES datasets, their materials coverage, and deposition rates is beyond the scope of this review. Nevertheless, the fragmentation is clear from both the available repositories and the ARPES ML literature. For example, embeddings pretrained on ImageNet using self-supervised learning do not cluster photoemission cuts effectively. This result supports the need for ARPES-specific image databases and pretext tasks designed for photoemission data~\cite{ekahana2023}. Similarly, labeled datasets for real-time experimental alignment remain scarce, and  researchers are increasingly using physics informed simulators and Sim2Real training to compensate for the lack of a shared experimental data~\cite{na2025}. Some very large  datasets have been released, including  multi-terabyte angle scan collections that were later used to train  ARPESNet-style autoencoders~\cite{zenodo_arpes_anglescan_2024,autoencoder2025}. However, these remain isolated deposits rather than parts of a universally maintained, multi task ARPES training resource.

% \begin{figure}[htbp]
% \centering
% \includegraphics[width=\textwidth]{fig06_data_models}
% \caption{Sporadic open ARPES data versus sharing pretrained task models: (a)~repository collage; (b)~MPES workflow schematic~\cite{xian2020mpes}; (c)~raw-data inertia versus model-hub concept  Draft collage; redraw/permission before submission.}
% \label{fig:data}
% \end{figure}

The individual datasets are still scientifically useful, and many of the tools needed to create a more reusable data landscape already exist. At EPFL, Infoscience catalogs publications~\cite{epfl_infoscience}, while related ARPES datasets may be hosted in Materials Cloud, as in the case of the  LACUS tr-ARPES dataset~\cite{materialscloud_arpes2017}. The Paul Scherrer Institute (PSI) similarly provides publication-linked ARPES data through its institutional data catalog and opendata.swiss~\cite{psi_opendata_schroter2020}.  Zenodo has also become a general purpose repository for raw and processed ARPES data, including tr-ARPES volumes~\cite{zenodo_arpes_wse2_2022} and large three dimensional angle scans that have already been reused for ARPESNet-style autoencoders~\cite{zenodo_arpes_anglescan_2024,autoencoder2025}.  On the data standard side, NeXus provides a common framework originally developed for X-ray, neutron, and muon experiments~\cite{nexus2015}.  Fairmat's NXmpes and ARPES application definitions, together with NOMAD and MPES end-to-end workflows, show how photoemission data can be stored with FAIR (findable, accessible, interoperable, reusable) metadata rather than as opaque binary files~\cite{nomad2023,xian2020mpes,fairmat_nxmpes2024}. Extending this, the DOE pathfinder Integrated Scientific Agentic AI for Catalysis (ISAAC)~\cite{doe_isaac_pathfinder} suggests archiving schema-validated knowledge rather than raw data. Therefore, the main bottleneck is not necessarily the lack of infrastructure, but its uneven adoption  across the ARPES community. Although we do not attempt to count every public ARPES dataset, the examples above show that useful training data already exist in different formats and with different levels of metadata. Facility roadmaps, such as the AI@ALS workshop report, also point to a second bottleneck related to data collection. Even when repositories and algorithms are available, scaling these approaches requires common application programming interfaces, data-streaming pathways, and control software. Without these tools, each beamline or endstation must develop a separate interface before the measured spectra can be incorporated into a machine-learning loop~\cite{alsworkshop2024}.

The questions of data scarcity and curation also lead to a more fundamental issue: what should be considered the ground truth in an ARPES database, particularly if the long-term goal is to develop an ARPES foundation model? For denoising, a long-acquisition map is often treated as the reference for a short-acquisition measurement, but surface aging, beam drift, sample charging, photon-flux changes, and related instabilities can make the two states differ, so a longer dwell is not automatically perfect ground truth. Broader criteria for validating denoised or reconstructed spectra---preservation of peak positions, linewidths, and related physical quantities, rather than visual smoothness alone---are discussed in Sec.~\ref{sec:denoising}. Synthetic datasets such as \textsc{aurelia} remain useful when experimental labels are scarce~\cite{na2025}, but should complement experimentally anchored labels rather than replace them. Expert-provided labels will remain important and should record the contributor, annotation procedure, confidence level, and physical criteria used. Another possibility is to reduce the dependence on manual labels by developing ARPES specific self-supervised learning tasks that include information about photoemission physics. For example, these tasks could use consistency between symmetry allowed momentum regions, neighboring energy momentum cuts, photon energy dependent scans, spectral function constraints, or self-energy relationships. In this case, the model can learn useful representations directly from the structure of the ARPES data instead of relying only on labels assigned by researchers~\cite{ekahana2023}.

Given the challenges of data scarcity, curation, and ground truth determination, a denser shared archive would certainly accelerate the development of an ARPES foundation model. However, creating a single, ImageNet-style database remains a highly demanding and logistically complex long term target across all aspects of adoption, management, and governance. For example, we note here that the upload inertia remains high: spectra are expensive to acquire and often establish the priority of discovery or new ideas. Meanwhile, the scientific enterprise mainly rewards papers, grants, and recognition, and not data curation labor. Although citation pathways for datasets already exist, like the digital object identifiers (DOIs) through DataCite and journal data sharing policies~\cite{datacite}, they do not by themselves create a dense, task labeled dataset if many groups do not fully adopt the initiative.

Here, we propose a more realistic aim which is a federated open archive with several small task specific benchmarks and models. In other words, we propose to share the pretrained task models, not as a substitute for bigger data deposits, but as a low inertia alternative path towards the adoption of AI in the ARPES pipeline, similar to how ISAAC prioritizes knowledge over raw data. As groups adopt ML locally, they can release weights together with model cards that state task, architecture, training domain (photon energy, analyzer class, materials family), known failure modes, and a fine tuning recipe. Concrete examples already surveyed in this review including the VAE denoiser~\cite{restrepo2022}, an SRCNN enhancer~\cite{peng2020}, SSL and graph-based domain labelers~\cite{ekahana2023,graphconv2025}, and theory-side weights such as DeepH- or SlaKoNet-style Hamiltonian surrogates and GW-correction models~\cite{deeph2022,slakonet2025,knosgaard2022}. Other groups could download these models, fine tune them using their own data, and upload the new versions together with information about the additional training domain and changes in performance. The repository could then track the development of each model, including its authors, training data, revisions, and version history. Over time, the models would evolve according to the needs and level of adoption within the community. However, this shared pretrained model proposal will only be useful if several practical issues are addressed:
\begin{enumerate}
  \item \emph{Portability.} A denoiser trained on soft-X-ray $\mu$ARPES at one synchrotron cannot be expected to work directly on data from a laboratory He-lamp system.  Fine-tuning or clearly documented tests of the domain shift will be required before the model can be reliably transferred to a different experimental setup.
  \item \emph{Versioning and citation;} Every model release should have a permanent identifier such as DOIs or versioned repository tags, together with a clear change log. If a published figure was generated using a particular model, the paper should identify the exact weights and the software version used.
  \item \emph{Stewardship.} Each model should have clearly identified maintainers who are responsible for documenting updates, responding to reported problems, releasing improved versions, and deprecating models that are no longer reliable. Initially, this responsibility will most likely remain with the university or national laboratory group that developed the model.
  \item \emph{Benchmarking and Limitations.} Users must be informed about both the performance and the limitation of the model. Small shared test sets should be developed for individual tasks such as denoising, domain labeling, band extraction, and bare band generation. The performance stated in the model card should be tested against these independent datasets. Without such testing, a model card may become mainly an advertisement and should not be treated as reliable information by an AI assistant or automated workflow.
  \item \emph{Training data documentation.} Releasing weights without any description of the training data greatly limits scientific reuse and reproducibility. This information is needed even if the full training data cannot be deposited publicly. The release should also describe all pre-processing steps \& code and a reference input output example. The model should be inherently verifiable. 
\end{enumerate}

 % including cropping, interpolation, background subtraction, intensity normalization, units, axis orientation, data reshaping 
Analysis frameworks that already track data provenance, such as PyARPES, and newer toolkits with optional ML components, such as peaks, are natural places to connect model cards and pretrained weights. Using existing analysis environments in this way may reduce the barrier to adopting  shared  models~\cite{pyarpes2020,peaks2025}.

Even with these requirements, the proposed model sharing approach may face an initial cold start problem. A practical starting point would be to repackage task specific models already published in the ARPES literature and at the same time, small benchmark datasets could be introduced through workshops, beamline initiatives, or facility led challenges. Instead of waiting for participation from the entire community, each task could begin with one carefully prepared public test set and an agreed-upon metric. These metrics could include reconstruction error relative to a longer acquisition, preservation of extracted physical quantities, or domain label accuracy. Journals, universities, and beamline facilities could further encourage this practice by asking authors to provide a model DOI together with the paper DOI whenever a machine-learning model is central to the reported result.

%\begin{figure}[htbp]
%  \centering
%  \includegraphics[width=0.9\textwidth]{image_3d3f9ca0.jpg}
%  \caption{\textbf{Cutesy version of 6 automation levels} The prompt used for \textbf{nanobanana lite 2} was ``could you generate a cutesy illustration of L0-L5? for example, the last one could be scientist kicking back and watching the experiemnt.  the first one could be the same scientist sweating over turning the knobs."}
%  \label{fig:cute6}
%\end{figure}

\section{Perspective}
\label{sec:perspective}

\subsection*{Towards the ``Self-driving ARPES Lab"}

Looking across the complete ARPES workflow, from sample preparation and measurement to data analysis and theoretical interpretation, AI is already being introduced at many different stages. However, these developments are at very different levels of maturity, applicability, and integrability. As discussed throughout this review, task-specific methods have been demonstrated for denoising, adaptive acquisition, domain labeling, band and self-energy extraction, and rapid electronic structure calculations. Most of these tools have so far been developed independently. The next important step is therefore to connect them so that information generated at one stage of an ARPES experiment can guide what happens next. A fully integrated AI framework for ARPES does not yet exist, but many of the individual pieces needed to build one are beginning to emerge.

To describe how ARPES could progress toward this goal, we adapt the six-level automation framework originally introduced for self driving vehicles through the SAE J3016 standard~\cite{saej3016_2021} and recently adapted by Le Houx as the {BASE} Scale (Benchmarking Autonomy in Scientific Experiments) for beamline automation~\cite{lehoux2026benchmarking}. As illustrated in Figure~\ref{fig:6automationlevel}, the framework structures the transition from fully manual operations to complete autonomy, based on where the decision-making authority resides:

\begin{figure}[htbp]
  \centering
  \includegraphics[width=0.9\textwidth]{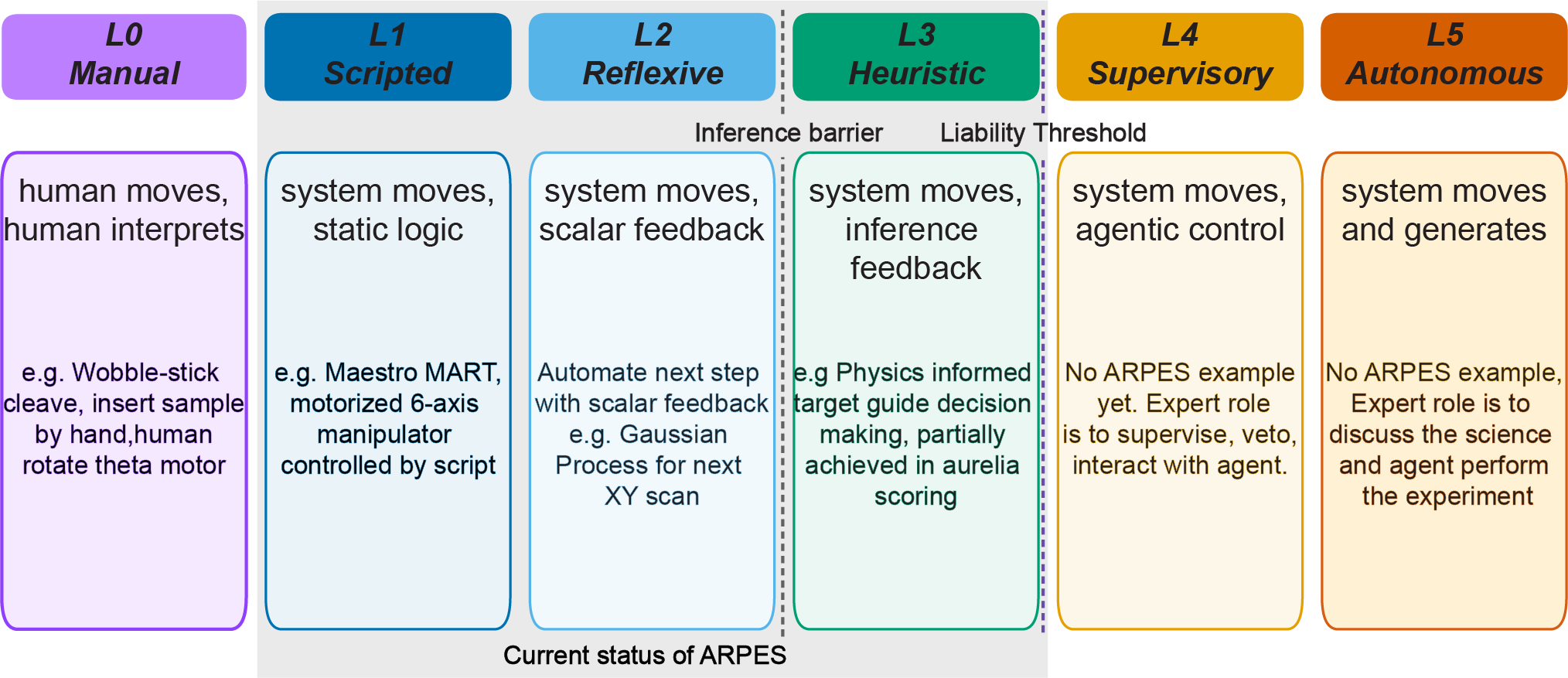}
  \caption{\textbf{The six automation levels of ARPES}, adapted from Ref.~\cite{lehoux2026benchmarking}. \textbf{Level 0 (Manual)} represents fully manual ARPES measurements with only the minimal necessary motorization. \textbf{Level 1 (Scripted)} represents the stage at which full motorization is realized and predefined scripts control all motion. \textbf{Level 2 (Reflexive)} represents the stage at which a subset of decisions is transferred to pre defined scalar feedback logic. \textbf{Level 3 (Heuristic)} represents the stage at which decisions are made heuristically through physics-informed inference. \textbf{Level 4 (Supervisory)} represents the stage at which an expert supervises an agentic bot controlling the experiment, while retaining key decision points and the right to veto the agent's actions. \textbf{Level 5 (Autonomous)} represents the stage at which the human engages with the agent at the level of scientific discussion, with minimal direct decision-making. Representative example is an agent that can propose and independently execute a new experiment based on prior discussion.}
  \label{fig:6automationlevel}
\end{figure}

\begin{itemize}
    \item \textbf{L0 (Manual):} This represents the traditional ARPES experiment, where most operations require direct expert involvement. The researcher transfers and prepares the sample, aligns it for measurement, selects the experimental conditions, and decides what to measure. Mechanical operations, e.g. wobble stick and transfer arm, are mostly manual.
    
    \item \textbf{L1 (Scripted):} At this level, mechanical operations are motorized and controlled through software. Sample movement, manipulator positioning, and predefined measurement sequences can be executed automatically, but the researcher still decides and performs preferred actions. This level is fully established in systems like the MART (MAESTRO Area Rapid Transit) architecture at the MAESTRO beamline at the ALS~\cite{maestrohighlight}.

    \item \textbf{L2 (Reflexive):} Here, individual experimental decisions begin to be made automatically using feedback from the measurement, as in the adaptive acquisition and domain-labeling examples of Sec.~\ref{sec:experiment}.
    
    \item \textbf{L3 (Heuristic):} At this stage, automated decisions begin to incorporate physics-informed criteria (spectral features, symmetry, quality metrics), as illustrated by the physics-parameterized approaches in Sec.~\ref{sec:experiment}. The system makes informed decisions, while the researcher continues to define the overall experimental strategy.
    
    \item \textbf{L4 (Supervisory):} The next step would be to connect several of these individual capabilities through an AI agent. This agent coordinates sample positioning, measurement, real-time analysis, and theoretical calculations. The researcher would define the scientific objective and closely oversee the experiment, while the agent handles operational detail. Importantly, the researcher would still be able to intervene or override the system as needed.

    \item \textbf{L5 (Autonomous):} At the highest level, the researcher defines the scientific objective, leaving the system with full autonomy. The AI agent plans the sequence of measurements, analyzes data in real-time, adjusts experimental conditions, compares results against theory, and determines the next measurement entirely on its own. Eventually, such systems could also coordinate measurements across complementary instruments. The ARI beamline concept at NSLS-II, for example, is designed to combine ARPES and RIXS in a common soft X-ray environment~\cite{he2020wave}, and such multi-probe facilities could later host coordinated autonomous workflows.
\end{itemize}

\subsection*{Current Status and Future Trajectories}

Present day ARPES does not fit neatly into any single one of these levels. Different beamlines have very different levels of automation, and even within the same endstation some operations may be highly automated while others remain completely manual. Overall, most current ARPES systems fall somewhere between \textbf{L1 and L3}. Similarly, the AI methods discussed in this review generally solve individual parts of the problem rather than controlling the complete experiment.

Moving beyond this stage requires overcoming two major challenges, which we illustrate in Figure~\ref{fig:6automationlevel}. The first is an inference barrier. Motorizing hardware and automating predefined scans are relatively straightforward compared with allowing a machine to interpret an ARPES spectrum and decide what should be measured next. Such decisions require understanding whether a spectral feature is physically meaningful, whether the measurement quality is sufficient, and whether changing parameters such as photon energy, polarization, temperature, or sample position will provide useful new information. Making these decisions reliably will require better training data, standardized metadata, real-time access to experimental data, uncertainty estimates, and closer integration between beamline-control and analysis software.

A second challenge appears when AI begins to directly control experimental hardware. At this point, trust and control become as important as model performance. An autonomous system should not only suggest an experimental action but should also operate within clearly defined limits, keep track of the decisions it makes, and allow the researcher to understand and override those decisions when necessary. This becomes particularly important when controlling expensive instrumentation or applying experimental conditions that could damage a sample or device.

At present, fully integrated \textbf{L4 or L5} ARPES experiments have not yet been demonstrated. However, many of the pieces required to move in this direction already exist independently. An important next step is therefore to develop shared and reusable models that can be transferred between experiments and beamlines. Building larger and better organized ARPES datasets will certainly help, but this does not necessarily mean that every future model must rely on a very large experimental training set. Physics informed simulations, synthetic data, transfer learning, and models that continue learning during an experiment may provide equally important routes, particularly for condensed matter systems where experimental data remain limited.

Ultimately, a “self-driving ARPES lab” should not simply mean an ARPES experiment that runs without a scientist. A more useful goal is an experiment in which sample handling, measurement, data analysis, and theoretical calculations communicate with each other in real-time. Routine decisions could then increasingly be handled by the automated system, while the researcher focuses on the scientific question, evaluates unexpected observations, and decides whether the emerging picture makes physical sense. In this way, AI shifts the researcher's role away from repetitive operations and toward the most critical aspect of the ARPES workflow, i.e., the physics.

\section*{Declarations}

\subsection*{Acknowledgments}
J.K.\ acknowledges financial support from the U.S.\ Department of Energy, Office of Science, Office of Basic Energy Sciences, through award No.~DE-SC002549 (postdoc support), NSF-CAREER Award under Grant No.~DMR-2339309 (graduate student support) and Army Research Office under Grant Number W911NF-26-1-A042 (graduate student support). This research used resources of the Advanced Light Source, a DOE Office of Science User Facility under contract No.~DE-AC02-05CH11231.

\subsection*{Competing interests}
The authors declare that they have no competing interests.

\subsection*{Ethics approval}
Not applicable.

\subsection*{Consent to participate}
Not applicable.

\subsection*{Consent for publication}
Not applicable.

\subsection*{Availability of data and material}
This is a review article. No new datasets were generated. Literature sources cited in the text are publicly available through the corresponding publishers or repositories.

\subsection*{Code availability}
Not applicable.

\subsection*{Authors' contributions}
J.K.\ initiated the project, supervised the work and edited the manuscript. S.A.E.\ led the literature survey, initial manuscript drafting, initial figures handling and drafting, finalizing manuscript writing, and overall revisions. A.T.\ synthesized recent developments in AI \& ML within electronic structure research, conceptualized \& created the figures, illustrations, prepared the first draft and overall revisions of the manuscript. P.S.\ contributed critical commentary that shaped the Data landscape and Perspective sections and revised the manuscript. A.B., C.J., and E.R.\ contributed scientific discussion, critical review and revisions of the manuscript, and facility/ARPES expertise. All authors read and approved the final manuscript.

\subsection*{Use of large language models}
LLMs were used under the authors' direction to organize and synthesize literature notes. The manuscript was written entirely by the authors. Sporadically, LLMs were used to improve overall clarity, grammar and formatting. All scientific claims, interpretations, citations, and final wording are the authors', who take full responsibility for the content. No LLM is listed as an author of this work, and generative AI was not used for ideation or illustration.

\bibliography{bibliography}

@misc{monark_foundry,
  author = {{MonArk Quantum Foundry}},
  title = {an NSF program that is jointly led by Montana State University and the University of Arkansas.},
  url = {https://www.monarkfoundry.org},
  organization = {National Science Foundation},
  year = {2023}
}

@article{mcculloch1943logical,
  title={A logical calculus of the ideas immanent in nervous activity},
  author = {McCulloch, Warren S and Pitts, Walter},
  journal={Bulletin of Mathematical Biophysics},
  volume={5},
  number={4},
  pages={115--133},
  year={1943},
  publisher={Springer},
  doi={10.1007/BF02478259}
}

@inproceedings{macqueen1967multivariate,
  title     = {Some methods for classification and analysis of multivariate observations},
  author = {MacQueen, J.},
  booktitle = {Proceedings of the 5th Berkeley Symposium on Mathematical Statistics and Probability},
  volume    = {1},
  pages     = {281--297},
  year      = {1967},
  organization = {University of California Press},
  url       = {https://projecteuclid.org/euclid.bsmsp/1200512992}
}

@book{sutton1998reinforcement,
  title     = {Reinforcement Learning: An Introduction},
  author = {Sutton, Richard S. and Barto, Andrew G.},
  year      = {1998},
  publisher = {MIT Press},
  address   = {Cambridge, MA}
}

@book{bezdek2013pattern,
  title     = {Pattern Recognition with Fuzzy Objective Function Algorithms},
  author    = {Bezdek, James C.},
  year      = {1981},
  publisher = {Plenum Press},
  address   = {New York},
  isbn      = {0-306-40671-3}
}

@article{jordan2015machine,
  title={Machine learning: Trends, perspectives, and prospects},
  author = {Jordan, Michael I and Mitchell, Tom M},
  journal={Science},
  volume={349},
  number={6245},
  pages={255--260},
  year={2015},
  publisher={American Association for the Advancement of Science},
  doi={10.1126/science.aaa8415}
}

@article{lecun2015deep,
  title={Deep learning},
  author = {LeCun, Yann and Bengio, Yoshua and Hinton, Geoffrey},
  journal={Nature},
  volume={521},
  number={7553},
  pages={436--444},
  year={2015},
  publisher={Nature Publishing Group UK London},
  doi={10.1038/nature14539}
}

@article{silver2016mastering,
  title={Mastering the game of Go with deep neural networks and tree search},
  author = {Silver, David and Huang, Aja and Maddison, Chris J. and Guez, Arthur and Sifre, Laurent and van den Driessche, George and Schrittwieser, Julian and Antonoglou, Ioannis and Panneershelvam, Veda and Lanctot, Marc and Dieleman, Sander and Grewe, Dominik and Nham, John and Kalchbrenner, Nal and Sutskever, Ilya and Lillicrap, Timothy and Leach, Madeleine and Kavukcuoglu, Koray and Graepel, Thore and Hassabis, Demis},
  journal={Nature},
  volume={529},
  number={7587},
  pages={484--489},
  year={2016},
  publisher={Nature Publishing Group UK London},
  doi={10.1038/nature16961}
}

@article{pedregosa2011scikit,
  title   = {Scikit-learn: Machine learning in {Python}},
  author = {Pedregosa, Fabian and Varoquaux, Ga{\"e}l and Gramfort, Alexandre and Michel, Vincent and Thirion, Bertrand and Grisel, Olivier and Blondel, Mathieu and Prettenhofer, Peter and Weiss, Ron and Dubourg, Vincent and others},
  journal = {Journal of Machine Learning Research},
  volume  = {12},
  pages   = {2825--2830},
  year    = {2011},
  publisher = {JMLR.org},
  url     = {https://jmlr.org/papers/v12/pedregosa11a.html}
}

@inproceedings{paszke2019pytorch,
  title     = {{PyTorch}: An Imperative Style, High-Performance Deep Learning Library},
  author = {Paszke, Adam and Gross, Sam and Massa, Francisco and Lerer, Adam and Bradbury, James and Chanan, Gregory and Killeen, Trevor and Lin, Zeming and Gimelshein, Natalia and Antiga, Luca and Desmaison, Alban and Kopf, Andreas and Yang, Edward and DeVito, Zachary and Raison, Martin and Tejani, Alykhan and Chilamkurthy, Sasank and Steiner, Benoit and Fang, Lu and Bai, Junjie and Chintala, Soumith},
  booktitle = {Advances in Neural Information Processing Systems},
  volume    = {32},
  pages     = {8026--8037},
  year      = {2019},
  eprint    = {1912.01703},
  archivePrefix = {arXiv},
  url       = {https://arxiv.org/abs/1912.01703}
}

@inproceedings{abadi2016tensorflow,
  title     = {{TensorFlow}: A system for large-scale machine learning},
  author = {Abadi, Mart{\'\i}n and Barham, Paul and Chen, Jianmin and Chen, Zhifeng and Davis, Andy and Dean, Jeffrey and Devin, Matthieu and Ghemawat, Sanjay and Irving, Geoffrey and Isard, Michael and others},
  booktitle = {12th USENIX Symposium on Operating Systems Design and Implementation (OSDI 16)},
  pages     = {265--283},
  year      = {2016},
  url       = {https://www.usenix.org/conference/osdi16/technical-sessions/presentation/abadi}
}

@article{jumper2021highly,
  title={Highly accurate protein structure prediction with AlphaFold},
  author = {Jumper, John and Evans, Richard and Pritzel, Alexander and Green, Tim and Figurnov, Michael and Ronneberger, Olaf and Tunyasuvunakool, Kathryn and Bates, Russ and {\v{Z}}{\'\i}dek, Augustin and Potapenko, Anna and others},
  journal={Nature},
  volume={596},
  number={7873},
  pages={583--589},
  year={2021},
  publisher={Nature Publishing Group},
  doi={10.1038/s41586-021-03819-2}
}

@article{lam2023learning,
  title={Learning skillful medium-range global weather forecasting},
  author = {Lam, Remi and Sanchez-Gonzalez, Alvaro and Willson, Matthew and Wirnsberger, Peter and Fortunato, Meire and Alet, Ferran and Ravuri, Suman and Ewalds, Timo and Eaton-Rosen, Zach and Hu, Weihua and Merose, Alexander and Hoyer, Stephan and Holland, George and Vinyals, Oriol and Stott, Jacklynn and Pritzel, Alexander and Mohamed, Shakir and Battaglia, Peter},
  journal={Science},
  volume={382},
  number={6677},
  pages={1416--1421},
  year={2023},
  publisher={American Association for the Advancement of Science},
  doi={10.1126/science.adi2336}
}

@article{degrave2022magnetic,
  title={Magnetic control of tokamak plasmas through deep reinforcement learning},
  author = {Degrave, Jonas and Felici, Federico and Buchli, Jonas and Neunert, Michael and Tracey, Brendan and Carpanese, Francesco and Ewalds, Timo and Hafner, Roland and Abdolmaleki, Abbas and de las Casas, Diego and Donner, Craig and Fritz, Leslie and Galperti, Cristian and Huber, Andrea and Keeling, James and Tsimpoukelli, Maria and Kay, Jackie and Merle, Antoine and Moret, Jean-Marc and Noury, Seb and Pesamosca, Federico and Pfau, David and Sauter, Olivier and Sommariva, Cristian and Coda, Stefano and Duval, Basil and Fasoli, Ambrogio and Kohli, Pushmeet and Kavukcuoglu, Koray and Hassabis, Demis and Riedmiller, Martin},
  journal={Nature},
  volume={602},
  number={7897},
  pages={414--419},
  year={2022},
  publisher={Nature Publishing Group},
  doi={10.1038/s41586-021-04301-9}
}

@article{stokes2020deep,
  title={A deep learning approach to antibiotic discovery},
  author = {Stokes, Jonathan M. and Yang, Kevin and Swanson, Kyle and Jin, Wengong and Cubillos-Ruiz, Andres and Donghia, Nina M. and MacNair, Craig R. and French, Shawn and Carfrae, Lindsey A. and Bloom-Ackermann, Zohar and Tran, Victoria M. and Chiappino-Pepe, Anush and Badran, Ahmed H. and Andrews, Ian W. and Chory, Emma J. and Church, George M. and Brown, Eric D. and Jaakkola, Tommi S. and Barzilay, Regina and Collins, James J.},
  journal={Cell},
  volume={180},
  number={4},
  pages={688--702},
  year={2020},
  publisher={Elsevier},
  doi={10.1016/j.cell.2020.01.021}
}

@article{wang2023scientific,
  title={Scientific discovery in the age of artificial intelligence},
  author = {Wang, Hanchen and Fu, Tianfan and Du, Yuanqi and Gao, Wenhao and Huang, Kexin and Liu, Ziming and Chandak, Payal and Liu, Shengchao and Van Katwyk, Peter and Deac, Andreea and Anandkumar, Anima and Bergen, Karianne and Gomes, Carla P. and Ho, Shirley and Kohli, Pushmeet and Lasenby, Joan and Leskovec, Jure and Liu, Tie-Yan and Manrai, Arjun and Marks, Debora and Ramsundar, Bharath and Song, Le and Sun, Jimeng and Tang, Jian and Veli{\v{c}}kovic, Petar and Welling, Max and Zhang, Linfeng and Coley, Connor W. and Bengio, Yoshua and Zitnik, Marinka},
  journal={Nature},
  volume={620},
  number={7972},
  pages={47--60},
  year={2023},
  publisher={Nature Publishing Group UK London},
  doi={10.1038/s41586-023-06221-2}
}

@article{carleo2019,
  author = {Carleo, Giuseppe and Cirac, Ignacio and Cranmer, Kyle and Daudet, Laurent and Schuld, Maria and Tishby, Naftali and Vogt-Maranto, Leslie and Zdeborov{\'a}, Lenka},
  title   = {Machine learning and the physical sciences},
  journal = {Reviews of Modern Physics},
  volume  = {91}, number = {4}, pages = {045002}, year = {2019},
  doi     = {10.1103/RevModPhys.91.045002}
}

@article{carleo2017,
  author = {Carleo, Giuseppe and Troyer, Matthias},
  title   = {Solving the quantum many-body problem with artificial neural networks},
  journal = {Science},
  volume  = {355}, number = {6325}, pages = {602--606}, year = {2017},
  doi     = {10.1126/science.aag2302}
}

@article{carrasquilla2017,
  author = {Carrasquilla, Juan and Melko, Roger G.},
  title   = {Machine learning phases of matter},
  journal = {Nature Physics},
  volume  = {13}, pages = {431--434}, year = {2017},
  doi     = {10.1038/nphys4035}
}

@article{kulik2022,
  author = {Kulik, H J and Hammerschmidt, T and Schmidt, J and Botti, S and Marques, M A L and Boley, M and Scheffler, M and Todorovic, M and Rinke, P and Oses, C and Smolyanyuk, A and Curtarolo, S and Tkatchenko, A and Bart{\'o}k, A P and Manzhos, S and Ihara, M and Carrington, T and Behler, J and Isayev, O and Veit, M and Grisafi, A and Nigam, J and Ceriotti, M and Sch{\"u}tt, K T and Westermayr, J and Gastegger, M and Maurer, R J and Kalita, B and Burke, K and Nagai, R and Akashi, R and Sugino, O and Hermann, J and No{\'e}, F and Pilati, S and Draxl, C and Kuban, M and Rigamonti, S and Scheidgen, M and Esters, M and Hicks, D and Toher, C and Balachandran, P V and Tamblyn, I and Whitelam, S and Bellinger, C and Ghiringhelli, L M},
  title   = {Roadmap on machine learning in electronic structure},
  journal = {Electronic Structure},
  volume  = {4}, number = {2}, pages = {023004}, year = {2022},
  doi     = {10.1088/2516-1075/ac572f}
}

@article{deeph2022,
  author = {Li, He and Wang, Zun and Zou, Nianlong and Ye, Meng and Xu, Runzhang and Gong, Xiaoxun and Duan, Wenhui and Xu, Yong},
  title   = {Deep-learning density functional theory {Hamiltonian} for efficient ab initio electronic-structure calculation},
  journal = {Nature Computational Science},
  volume  = {2}, pages = {367--377}, year = {2022},
  doi     = {10.1038/s43588-022-00265-6}
}

@article{benmahmoud2020dos,
  author = {Ben Mahmoud, Chiheb and Anelli, Andrea and Cs{\'a}nyi, G{\'a}bor and Ceriotti, Michele},
  title   = {Learning the electronic density of states in condensed matter},
  journal = {Physical Review B},
  volume  = {102}, number = {23}, pages = {235130}, year = {2020},
  doi     = {10.1103/PhysRevB.102.235130}
}

@article{chandrasekaran2019,
  author = {Chandrasekaran, Anand and Kamal, Deepak and Batra, Rohit and Kim, Chiho and Chen, Lihua and Ramprasad, Rampi},
  title   = {Solving the electronic structure problem with machine learning},
  journal = {npj Computational Materials},
  volume  = {5}, pages = {22}, year = {2019},
  doi     = {10.1038/s41524-019-0162-7}
}

@article{schnorb2019,
  author = {Sch{\"u}tt, K. T. and Gastegger, M. and Tkatchenko, A. and M{\"u}ller, K.-R. and Maurer, R. J.},
  title   = {Unifying machine learning and quantum chemistry with a deep neural network for molecular wavefunctions},
  journal = {Nature Communications},
  volume  = {10}, pages = {5024}, year = {2019},
  doi     = {10.1038/s41467-019-12875-2}
}

@article{how2025petmaddos,
  author  = {How, Wei Bin and Febrer, Pol and Chong, Sanggyu and Mazitov, Arslan and Bigi, Filippo and Kellner, Matthias and Pozdnyakov, Sergey and Ceriotti, Michele},
  title   = {A universal machine learning model for the electronic density of states},
  journal = {Digital Discovery},
  volume  = {5},
  number  = {4},
  pages   = {1635--1649},
  year    = {2026},
  doi     = {10.1039/D5DD00557D}
}

@article{lou2026moire,
  author  = {Lou, Zekun and Lewis, Alan M. and Rossi, Mariana},
  title   = {Long-Range Machine Learning of Electron Density for Twisted Bilayer Moir{\'e} Materials},
  journal = {PRX Intelligence},
  volume  = {1},
  number  = {1},
  year    = {2026},
  doi     = {10.1103/4575-9cmx},
  eprint  = {2602.09938},
  archivePrefix = {arXiv},
  primaryClass  = {cond-mat.mtrl-sci}
}

@article{deephe3_2023,
  author = {Gong, Xiaoxun and Li, He and Zou, Nianlong and Xu, Runzhang and Duan, Wenhui and Xu, Yong},
  title   = {General framework for {E(3)}-equivariant neural network representation of density functional theory {Hamiltonian}},
  journal = {Nature Communications},
  volume  = {14}, pages = {2848}, year = {2023},
  doi     = {10.1038/s41467-023-38468-8}
}

@article{deephuniversal2024,
  author = {Wang, Yuxiang and Li, Yang and Tang, Zechen and Li, He and Yuan, Zilong and Tao, Honggeng and Zou, Nianlong and Bao, Ting and Liang, Xinghao and Chen, Zezhou and Xu, Shanghua and Bian, Ce and Xu, Zhiming and Wang, Chong and Si, Chen and Duan, Wenhui and Xu, Yong},
  title   = {Universal materials model of deep-learning density functional theory {Hamiltonian}},
  journal = {Science Bulletin},
  year    = {2024},
  volume={69},
  number={16},
  pages={2514--2521},
  doi     = {10.1016/j.scib.2024.06.011}
}

@misc{maestro,
  author = {{Advanced Light Source}},
  title        = {{MAESTRO}: Microscopic and Electronic Structure Observatory (beamline 7.0.2) --- automated {MART} sample transfer, {\textmu}ARPES, {PEEM} and nanoARPES},
  howpublished = {Advanced Light Source, Lawrence Berkeley National Laboratory},
  year         = {2016},
  note         = {\url{https://als.lbl.gov/beamlines/7-0-2/}}
}

@misc{maestrohighlight,
  author = {{Advanced Light Source}},
  title        = {{MAESTRO} Beamline Set to Open to Users},
  howpublished = {ALS Science Highlight, Lawrence Berkeley National Laboratory},
  year         = {2016},
  note         = {\url{https://als.lbl.gov/maestro-beamline-set-open-users/}}
}

@article{melton2020,
  author = {Melton, Charles N. and Noack, Marcus M. and Ohta, Taisuke and Beechem, Thomas E. and Robinson, Jeremy and Zhang, Xiaotian and Bostwick, Aaron and Jozwiak, Chris and Koch, Roland J. and Zwart, Petrus H. and Hexemer, Alexander and Rotenberg, Eli},
  title   = {K-means-driven {Gaussian} process data collection for angle-resolved photoemission spectroscopy},
  journal = {Machine Learning: Science and Technology},
  volume  = {1}, number = {4}, pages = {045015}, year = {2020},
  doi     = {10.1088/2632-2153/abab61}
}

@article{masubuchi2018,
  author = {Masubuchi, Satoru and Morimoto, Masataka and Morikawa, Sei and Onodera, Momoko and Asakawa, Yuta and Watanabe, Kenji and Taniguchi, Takashi and Machida, Tomoki},
  title   = {Autonomous robotic searching and assembly of two-dimensional crystals to build van der {Waals} superlattices},
  journal = {Nature Communications},
  volume  = {9}, pages = {1413}, year = {2018},
  doi     = {10.1038/s41467-018-03723-w}
}

@article{flakeseg2025,
  author = {Li, Yutao and Sherlock, Logan and Benderson, Ryan and Ostrom, Daniel and Chen, Huandong and Fujita, Kazuhiro and Pasupathy, Abhay},
  title         = {A practical flake segmentation and indexing pipeline for automated {2D} material stacking},
  journal       = {arXiv preprint},
  year          = {2025},
  eprint        = {2509.01826},
  archivePrefix = {arXiv},
  url           = {https://arxiv.org/abs/2509.01826}
}

@article{restrepo2022,
  author = {Restrepo, Francisco and Zhao, Junjing and Chatterjee, Utpal},
  title   = {Denoising and feature extraction in photoemission spectra with variational auto-encoder neural networks},
  journal = {Review of Scientific Instruments},
  volume  = {93}, number = {6}, pages = {065106}, year = {2022},
  doi     = {10.1063/5.0090051}
}

@article{kim2021,
  author = {Kim, Younsik and Oh, Dongjin and Huh, Soonsang and Song, Dongjoon and Jeong, Sunbeom and Kwon, Junyoung and Kim, Minsoo and Kim, Donghan and Ryu, Hanyoung and Jung, Jongkeun and Kyung, Wonshik and Sohn, Byungmin and Lee, Suyoung and Hyun, Jounghoon and Lee, Yeonghoon and Kim, Yeongkwan and Kim, Changyoung},
  title   = {Deep learning-based statistical noise reduction for multidimensional spectral data},
  journal = {Review of Scientific Instruments},
  volume  = {92}, number = {7}, pages = {073901}, year = {2021},
  doi     = {10.1063/5.0054920}
}

@article{prb2026,
  author = {Cheng, Gongyi and Guo, Ziyang and Yan, Zhengcong and Ma, Xiaochuan and Qiu, Yichen and Shen, Guangzhen and Li, Xintong and Tan, Shijing and Wang, Bing},
  title   = {Deep learning approach for {ARPES} and time-resolved {ARPES}: gradient attention {U-net} for denoising and reconstruction},
  journal = {Physical Review B},
  volume  = {114}, number = {5}, pages = {055414}, year = {2026},
  doi     = {10.1103/53wd-bg7k}
}

@article{peng2020,
  author = {Peng, Han and Gao, Xiang and He, Yu and Li, Yiwei and Ji, Yuchen and Liu, Chuhang and Ekahana, Sandy A. and Pei, Ding and Liu, Zhongkai and Shen, Zhixun and Chen, Yulin},
  title   = {Super resolution convolutional neural network for feature extraction in spectroscopic data},
  journal = {Review of Scientific Instruments},
  volume  = {91}, number = {3}, pages = {033905}, year = {2020},
  doi     = {10.1063/1.5132586}
}

@article{ekahana2023,
  author = {Ekahana, Sandy Adhitia and Winata, Genta Indra and Soh, Yona and Tamai, Anna and Milan, Radovic and Aeppli, Gabriel and Shi, Ming},
  title   = {Transfer learning application of self-supervised learning in {ARPES}},
  journal = {Machine Learning: Science and Technology},
  volume  = {4}, number = {3}, pages = {035021}, year = {2023},
  doi     = {10.1088/2632-2153/aced7d}
}

@article{mahlow2023model,
  title={Model-independent quantum phases classifier},
  author={Mahlow, Felipe and Luiz, Fabr{\'\i}cio Souza and Malvezzi, Andr{\'e} Luiz and Fanchini, Felipe Fernandes},
  journal={Scientific Reports},
  volume={13},
  number={1},
  pages={14411},
  year={2023},
  publisher={Nature Publishing Group UK London}
}

@article{nop2025faithful,
  title={Faithful novel machine learning for predicting quantum properties},
  author={Nop, Gavin and Mundy, Micah and Smith, Jonathan DH and Paudyal, Durga},
  journal={npj Computational Materials},
  volume={11},
  number={1},
  pages={244},
  year={2025},
  publisher={Nature Publishing Group UK London}
}

@article{samarakoon2022integration,
  title={Integration of machine learning with neutron scattering for the Hamiltonian tuning of spin ice under pressure},
  author={Samarakoon, Anjana and Tennant, D Alan and Ye, Feng and Zhang, Qiang and Grigera, Santiago A},
  journal={Communications Materials},
  volume={3},
  number={1},
  pages={84},
  year={2022},
  publisher={Nature Publishing Group UK London}
}

@article{kazeev2023sparse,
  title={Sparse representation for machine learning the properties of defects in 2D materials},
  author={Kazeev, Nikita and Al-Maeeni, Abdalaziz Rashid and Romanov, Ignat and Faleev, Maxim and Lukin, Ruslan and Tormasov, Alexander and Castro Neto, AH and Novoselov, Kostya S and Huang, Pengru and Ustyuzhanin, Andrey},
  journal={npj Computational Materials},
  volume={9},
  number={1},
  pages={113},
  year={2023},
  publisher={Nature Publishing Group UK London}
}

@article{graphconv2025,
  author  = {Santoso Sugiarto, Hendrik and Adhitia Ekahana, Sandy and Christofer Wijaya, Bryan and Indra Winata, Genta and Soh, Y and Aeppli, G},
  title   = {Enhancing transfer learning in angle-resolved photoemission spectroscopy ({ARPES}) with spatially-aware representations via graph convolution},
  journal = {Machine Learning: Science and Technology},
  year    = {2026},
  volume  = {7},
  number  = {2},
  pages   = {025007},
  doi     = {10.1088/2632-2153/ae3fe7}
}

@article{xian2023,
  author = {Xian, R. Patrick and Stimper, Vincent and Zacharias, Marios and Dendzik, Maciej and Dong, Shuo and Beaulieu, Samuel and Sch{\"o}lkopf, Bernhard and Wolf, Martin and Rettig, Laurenz and Carbogno, Christian and Bauer, Stefan and Ernstorfer, Ralph},
  title   = {A machine learning route between band mapping and band structure},
  journal = {Nature Computational Science},
  volume  = {3}, number = {1}, pages = {101--114}, year = {2023},
  doi     = {10.1038/s43588-022-00382-2}
}

@article{autoencoder2025,
  author = {{\'A}g\'ustsson, Steinn \'Ymir and Haque, Mohammad Ahsanul and Truong, Thi Tam and others},
  title   = {An autoencoder for compressing angle-resolved photoemission spectroscopy data},
  journal = {Machine Learning: Science and Technology},
  volume  = {6},
  number  = {1},
  pages   = {015014},
  year    = {2025},
  doi     = {10.1088/2632-2153/ada8f2}
}

@article{msca2024,
  author = {Bian, Lingzhu and Liu, Chen and Zhang, Zhen and Huang, Yingke and Pan, Xinyu and Zhang, Yi and Wang, Jiaou and Dudin, Pavel and Avila, Jose and Chen, Zhesheng and Dong, Yuhui},
  title   = {Automatic extraction of fine structural information in angle-resolved photoemission spectroscopy by multi-stage clustering algorithm},
  journal = {Communications Physics},
  volume  = {7}, pages = {379}, year = {2024},
  doi     = {10.1038/s42005-024-01878-1}
}

@article{chen2025dann,
  author = {Chen, Xu and Sun, Yuanjie and Hruska, Eugen and Dixit, Vivek and Yang, Jinming and He, Yu and Wang, Yao and Liu, Fang},
  title   = {Detecting thermodynamic phase transition via explainable machine learning of photoemission spectroscopy},
  journal = {Newton},
  volume  = {1}, number = {3}, pages = {100066}, year = {2025},
  doi     = {10.1016/j.newton.2025.100066}
}

@article{chitturi2023,
  author = {Chitturi, Sathya R. and Ji, Zhurun and Petsch, Alexander N. and Peng, Cheng and Chen, Zhantao and Plumley, Rajan and Dunne, Mike and Mardanya, Sougata and Chowdhury, Sugata and Chen, Hongwei and Bansil, Arun and Feiguin, Adrian and Kolesnikov, Alexander I. and Prabhakaran, Dharmalingam and Hayden, Stephen M. and Ratner, Daniel and Jia, Chunjing and Nashed, Youssef and Turner, Joshua J.},
  title   = {Capturing dynamical correlations using implicit neural representations},
  journal = {Nature Communications},
  volume  = {14}, pages = {5852}, year = {2023},
  doi     = {10.1038/s41467-023-41378-4}
}

@article{kalinin2023,
  author = {Kalinin, Sergei V. and Mukherjee, Debangshu and Roccapriore, Kevin and Blaiszik, Benjamin J. and Ghosh, Ayana and Ziatdinov, Maxim A. and Al-Najjar, Anees and Doty, Christina and Akers, Sarah and Rao, Nageswara S. and Agar, Joshua C. and Spurgeon, Steven R.},
  title   = {Machine learning for automated experimentation in scanning transmission electron microscopy},
  journal = {npj Computational Materials},
  volume  = {9}, pages = {227}, year = {2023},
  doi     = {10.1038/s41524-023-01142-0}
}

@article{pithan2023,
  author = {Pithan, Linus and Starostin, Vladimir and Mare{\v{c}}ek, David and Petersdorf, Lukas and V{\"o}lter, Constantin and Munteanu, Valentin and Jankowski, Maciej and Konovalov, Oleg and Gerlach, Alexander and Hinderhofer, Alexander and Murphy, Bridget and Kowarik, Stefan and Schreiber, Frank},
  title   = {Closing the loop: autonomous experiments enabled by machine-learning-based online data analysis in synchrotron beamline environments},
  journal = {Journal of Synchrotron Radiation},
  volume  = {30}, number = {6}, pages = {1064--1075}, year = {2023},
  doi     = {10.1107/S160057752300749X}
}

@article{aims2026,
  author = {Qiu, Siyuan and Suh, Philip D. and Tran, Nhat Huy and Wang, Xirui and Park, Heonjoon and Akin, Kutay and Multani, Kevin K. S. and Jung, Seungwon and Cai, Wenkai and Liu, Xinyu and Zhu, Ziyan and Jia, Chunjing and Chen, Zhantao and Shen, Zhixun and Ji, Zhurun},
  title   = {{AIMS}: an uncertainty-aware {AI} experimentalist for quantum matter},
  journal = {arXiv preprint},
  year    = {2026}, eprint = {2607.16544}, archivePrefix = {arXiv},
  url     = {https://arxiv.org/abs/2607.16544}
}

@article{heteroassembly2026,
  author = {Li, Xiaoxi and He, Jinkun and Liu, Haojie and Liu, Xipeng and Wu, Zewen and Li, Jing and Zhao, Kai and Li, Shan and Sun, Xingdan and Fan, Xiaoxue and Xiong, Zhiren and Wu, Xingguang and Sha, Xuanzhe and Lin, Zhili and Yang, Caixia and Han, Luosha and Xu, Jie and Pei, Woye and Yang, Kaining and Zhang, Jing and Feng, Xiaolong and Zhang, Tongyao and Liang, Zhu and Watanabe, Kenji and Taniguchi, Takashi and Tian, Ming and Wan, Neng and Zhang, Jing and Lu, Jianming and Hong, Wenjing and Han, Zheng Vitto},
  title         = {An {AI}-driven robotic system for two-dimensional hetero-assemblies},
  journal       = {arXiv preprint},
  year          = {2026},
  eprint        = {2605.20420},
  archivePrefix = {arXiv},
  url           = {https://arxiv.org/abs/2605.20420}
}

@article{qumus2026,
  author = {Shi, Lihan and Zheng, Zhaoyi Joy and Juan, Xinzhe and Wang, Yimin and Yin, Ming and Sengupta, Mayank and Wolinski, Kristina and Jia, Yanyu and Shi, Jingzhi and Saucedo, Derek and Saggi, Neill and Guan, Haosen and Watanabe, Kenji and Taniguchi, Takashi and Yazdani, Ali and Wang, Mengdi and Wu, Sanfeng},
  title         = {Qumus: Realization of an embodied {AI} quantum material experimentalist},
  journal       = {arXiv preprint},
  year          = {2026},
  eprint        = {2605.18407},
  archivePrefix = {arXiv},
  url           = {https://arxiv.org/abs/2605.18407}
}

@article{cheng2026materials,
  author = {Cheng, Mouyang and Fu, Chu-Liang and Okabe, Ryotaro and Chotrattanapituk, Abhijatmedhi and Boonkird, Artittaya and Hung, Nguyen Tuan and Li, Mingda},
  title   = {Artificial intelligence-driven approaches for materials design and discovery},
  journal = {Nature Materials},
  year    = {2026},
  volume={25},
  number={2},
  pages={174--190},
  doi     = {10.1038/s41563-025-02403-7},
  note    = {OSTI ID: 3030140}
}

@article{szymanski2023,
  author = {Szymanski, Nathan J. and Rendy, Bernardus and Fei, Yuxing and Kumar, Rishi E. and He, Tanjin and Milsted, David and McDermott, Matthew J. and Gallant, Max and Cubuk, Ekin Dogus and Merchant, Amil and Kim, Haegyeom and Jain, Anubhav and Bartel, Christopher J. and Persson, Kristin and Zeng, Yan and Ceder, Gerbrand},
  title   = {An autonomous laboratory for the accelerated synthesis of inorganic materials},
  journal = {Nature},
  volume  = {624},
  pages   = {86--91},
  year    = {2023},
  doi     = {10.1038/s41586-023-06734-w}
}

@article{xian2020mpes,
  author = {Xian, R. Patrick and Rettig, Laurenz and Ernstorfer, Ralph},
  title   = {An open-source, end-to-end workflow for multidimensional photoemission spectroscopy},
  journal = {Scientific Data},
  volume  = {7},
  pages   = {442},
  year    = {2020},
  doi     = {10.1038/s41597-020-00769-8}
}

@misc{materialscloud_arpes2017,
  author = {Crepaldi, Alberto and Roth, Silvan and Gatti, Gianmarco and Arrell, Christopher A. and Ojeda, Jos{\'e} and van Mourik, Frank and Bugnon, Philippe and Magrez, Arnaud and Berger, Helmuth and Chergui, Majed and Grioni, Marco},
  title        = {Time-resolved {ARPES} at {LACUS}: band structure and ultrafast electron dynamics of solids (Materials Cloud Archive)},
  year         = {2021},
  howpublished = {Materials Cloud},
  doi          = {10.24435/materialscloud:jy-5c},
    note         = {Dataset linked to EPFL/LACUS tr-ARPES work; Swiss open materials data}
}

@misc{psi_opendata_schroter2020,
  author = {Schr{\"o}ter, Niels Bernhard},
  title        = {{ARPES} data linked to the publication {N.B.M.} Schr{\"o}ter et al., Science aaz3480 (2020)},
  year         = {2020},
  howpublished = {{PSI} Data Catalog / opendata.swiss},
  doi          = {10.16907/D4DF1E49-6A62-40B5-B343-A8BE757521F2},
  url          = {https://opendata.swiss/en/dataset/arpes-data-linked-to-the-publication-n-b-m-schroter-et-al-science-aaz3480-20201}
}

@misc{zenodo_arpes_wse2_2022,
  author = {Maklar, J. and Dong, S. and Beaulieu, S. and Pincelli, T. and Dendzik, M. and Windsor, Y. W. and Xian, R. P. and Wolf, M. and Ernstorfer, R. and Rettig, L.},
  title        = {Time-resolved {ARPES} {RAW} data of bulk {WSe$_2$} (momentum microscope)},
  year         = {2022},
  howpublished = {Zenodo},
  doi          = {10.5281/zenodo.6369728}
}

@misc{zenodo_arpes_anglescan_2024,
  author = {{\'A}g{\'u}stsson, Steinn {\'Y}mir and Bianchi, Marco and Hofmann, Philip},
  title        = {{3D ARPES} angle scan collection},
  year         = {2024},
  howpublished = {Zenodo},
  doi          = {10.5281/zenodo.12665275},
    note         = {Used for ARPESNet autoencoder training}
}

@article{nomad2023,
  author = {Scheidgen, Markus and Himanen, Lauri and Ladines, Alvin Noe and Sikter, David and Nakhaee, Mohammad and Fekete, {\'A}d{\'a}m and Chang, Theodore and Golparvar, Amir and M{\'a}rquez, Jos{\'e} A. and Brockhauser, Sandor and Br{\"u}ckner, Sebastian and Ghiringhelli, Luca M. and Dietrich, Felix and Lehmberg, Daniel and Denell, Thea and Albino, Andrea and N{\"a}sstr{\"o}m, Hampus and Shabih, Sherjeel and Dobener, Florian and K{\"u}hbach, Markus and Mozumder, Rubel and Rudzinski, Joseph F. and Daelman, Nathan and Pizarro, Jos{\'e} M. and Kuban, Martin and Salazar, Cuauhtemoc and Ondra{\v{c}}ka, Pavel and Bungartz, Hans-Joachim and Draxl, Claudia},
  title   = {{NOMAD}: A distributed web-based platform for managing materials science research data},
  journal = {Journal of Open Source Software},
  volume  = {8},
  number  = {90},
  pages   = {5388},
  year    = {2023},
  doi     = {10.21105/joss.05388}
}

@article{nexus2015,
  author = {K{\"o}nnecke, Mark and Akeroyd, Frederick A. and Bernstein, Herbert J. and Brewster, Aaron S. and Campbell, Stuart I. and Clausen, Bj{\"o}rn and Cottrell, Stephen and Hoffmann, Jens Uwe and Jemian, Pete R. and M{\"a}nnicke, David and Osborn, Raymond and Peterson, Peter F. and Richter, Tobias and Suzuki, Jiro and Watts, Benjamin and Wintersberger, Eugen and Wuttke, Joachim},
  title   = {The {NeXus} data format},
  journal = {Journal of Applied Crystallography},
  volume  = {48},
  number  = {1},
  pages   = {301--305},
  year    = {2015},
  doi     = {10.1107/S1600576714027575}
}

@misc{datacite,
  author = {{DataCite}},
  title        = {{DataCite}: Connecting research, enabling discovery},
  year         = {2026},
  howpublished = {DataCite Consortium},
  url          = {https://datacite.org/},
  note         = {Persistent identifiers (DOIs) and citation infrastructure for research data and other outputs}
}

@misc{fairmat_nxmpes2024,
  author = {Pielsticker, Lukas and Dobener, Florian and Arora, Abeer and Pincelli, Tommaso and Weber, Heiko and Aeschlimann, Martin and Brockhauser, Sandor and Rettig, Laurenz},
  title        = {Collecting and storing {FAIR} (meta-)data: An {ARPES} example ({NXmpes} / {NOMAD})},
  year         = {2024},
  howpublished = {Zenodo},
  doi          = {10.5281/zenodo.14065847}
}

@misc{epfl_infoscience,
  author = {{EPFL}},
  title        = {{Infoscience}: {EPFL} institutional repository (publications and linked research outputs)},
  year         = {2026},
  howpublished = {{EPFL} open science infrastructure},
  url          = {https://infoscience.epfl.ch/},
  note         = {Swiss open-science institutional layer; datasets often deposited in Materials Cloud / Zenodo and linked from Infoscience}
}

@article{damascelli2003,
  author  = {Damascelli, Andrea and Hussain, Zahid and Shen, Zhi-Xun},
  title   = {Angle-resolved photoemission studies of the cuprate superconductors},
  journal = {Reviews of Modern Physics},
  volume  = {75},
  number  = {2},
  pages   = {473--541},
  year    = {2003},
  doi     = {10.1103/RevModPhys.75.473}
}

@article{cattelan2018perspective,
  title={A perspective on the application of spatially resolved ARPES for 2D materials},
  author = {Cattelan, Mattia and Fox, Neil A},
  journal={Nanomaterials},
  volume={8},
  number={5},
  pages={284},
  year={2018},
  publisher={MDPI},
  doi={10.3390/nano8050284}
}

@article{lv2019angle,
  title={Angle-resolved photoemission spectroscopy and its application to topological materials},
  author = {Lv, Baiqing and Qian, Tian and Ding, Hong},
  journal={Nature Reviews Physics},
  volume={1},
  number={10},
  pages={609--626},
  year={2019},
  publisher={Nature Publishing Group UK London},
  doi={10.1038/s42254-019-0088-5}
}

@article{hofmann2021,
  author = {Hofmann, Philip},
  title   = {Accessing the spectral function of in operando devices by angle-resolved photoemission spectroscopy},
  journal = {AVS Quantum Science},
  volume  = {3}, number = {2}, pages = {021101}, year = {2021},
  doi     = {10.1116/5.0038637}
}

@article{sobota2021,
  author = {Sobota, Jonathan A. and He, Yu and Shen, Zhi-Xun},
  title   = {Angle-resolved photoemission studies of quantum materials},
  journal = {Reviews of Modern Physics},
  volume  = {93}, number = {2}, pages = {025006}, year = {2021},
  doi     = {10.1103/RevModPhys.93.025006}
}

@article{zhang2022angle,
  title={Angle-resolved photoemission spectroscopy},
  author = {Zhang, Hongyun and Pincelli, Tommaso and Jozwiak, Chris and Kondo, Takeshi and Ernstorfer, Ralph and Sato, Takafumi and Zhou, Shuyun},
  journal={Nature Reviews Methods Primers},
  volume={2},
  number={1},
  pages={54},
  year={2022},
  publisher={Nature Publishing Group UK London},
  doi={10.1038/s43586-022-00133-7}
}

@article{moser2017experimentalist,
  title={An experimentalist's guide to the matrix element in angle resolved photoemission},
  author = {Moser, Simon},
  journal={Journal of Electron Spectroscopy and Related Phenomena},
  volume={214},
  pages={29--52},
  year={2017},
  publisher={Elsevier},
  doi={10.1016/j.elspec.2016.11.007}
}

@article{iwasawa2022,
  author = {Iwasawa, Hideaki and Ueno, Tetsuro and Masui, Takahiko and Tajima, Setsuko},
  title   = {Unsupervised clustering for identifying spatial inhomogeneity on local electronic structures},
  journal = {npj Quantum Materials},
  volume  = {7}, pages = {24}, year = {2022},
  doi     = {10.1038/s41535-021-00407-5}
}

@article{imamura2024,
  author = {Imamura, Masaki and Takahashi, Kazutoshi},
  title   = {Unsupervised learning of spatially-resolved {ARPES} spectra for epitaxially grown graphene via non-negative matrix factorization},
  journal = {Scientific Reports},
  volume  = {14}, pages = {24200}, year = {2024},
  doi     = {10.1038/s41598-024-73795-w}
}

@article{agustsson2024,
  author = {{\'A}g\'ustsson, Steinn \'Ymir and Jones, Alfred J. H. and Curcio, Davide and Ulstrup, S{\o}ren and Miwa, Jill and Mottin, Davide and Karras, Panagiotis and Hofmann, Philip},
  title   = {Autonomous micro-focus angle-resolved photoemission spectroscopy},
  journal = {Review of Scientific Instruments},
  volume  = {95}, number = {5}, pages = {055106}, year = {2024},
  doi     = {10.1063/5.0204663}
}

@article{iwasawaspin2024,
  author = {Iwasawa, Hideaki and Ueno, Tetsuro and Iwata, Takuma and Kuroda, Kenta and Kokh, Konstantin A. and Tereshchenko, Oleg E. and Miyamoto, Koji and Kimura, Akio and Okuda, Taichi},
  title   = {Efficiency improvement of spin-resolved {ARPES} experiments using {Gaussian} process regression},
  journal = {Scientific Reports},
  volume  = {14}, pages = {20970}, year = {2024},
  doi     = {10.1038/s41598-024-66704-8}
}

@article{pyarpes2020,
  author = {Stansbury, Conrad and Lanzara, Alessandra},
  title   = {{PyARPES}: an analysis framework for multimodal angle-resolved photoemission spectroscopies},
  journal = {SoftwareX},
  volume  = {12}, pages = {100472}, year = {2020},
  doi     = {10.1016/j.softx.2020.100472}
}

@article{peaks2025,
  author = {King, Phil D. C. and Edwards, Brendan and Mo, Shu and Antonelli, Tommaso and Morales, Edgar Abarca and Hart, Lewis and Trzaska, Liam},
  title   = {peaks: a {Python} package for analysis of angle-resolved photoemission and related spectroscopies},
  journal = {arXiv preprint},
  year    = {2025}, eprint = {2508.04803}, archivePrefix = {arXiv},
  url     = {https://arxiv.org/abs/2508.04803}
}

@article{na2025,
  author = {Na, MengXing and Zhou, Chris and Dufresne, Sydney K. Y. and Michiardi, Matteo and Damascelli, Andrea},
  title   = {A simulation-based training framework for machine-learning applications in {ARPES}},
  journal = {arXiv preprint},
  year    = {2025}, eprint = {2508.15983}, archivePrefix = {arXiv},
  url     = {https://arxiv.org/abs/2508.15983},
  note    = {Introduces the open-source synthetic ARPES simulator \textsc{aurelia}}
}

@article{alsworkshop2024,
  author = {Parkinson, Dilworth Y. and Chavez, Tanny and Choudhary, Monika and English, Damon and Hao, Guanhua and Hellert, Thorsten and Leemann, Simon C. and Nemsak, Slavomir and Rotenberg, Eli and Taylor, Andrea L. and Scholl, Andreas and White, Ashley A. and Islegen-Wojdyla, Antoine and Zwart, Petrus H. and Hexemer, Alexander},
  title   = {{AI@ALS} workshop report: machine learning needs at the Advanced Light Source},
  journal = {Synchrotron Radiation News},
  year    = {2024},
  volume={37},
  number={4},
  pages={49--64},
  doi     = {10.1080/08940886.2024.2391258}
}

@article{liu2023,
  author = {Liu, Junde and Huang, Dongchen and Yang, Yi-feng and Qian, Tian},
  title   = {Removing grid structure in angle-resolved photoemission spectra via deep learning method},
  journal = {Physical Review B},
  volume  = {107}, number = {16}, pages = {165106}, year = {2023},
  doi     = {10.1103/PhysRevB.107.165106}
}

@article{borisenko2026,
  author = {Borisenko, Sergey},
  title   = {The role of {A} in {ARPES}},
  journal = {Journal of Synchrotron Radiation},
  volume  = {33},
  number  = {3},
  pages   = {693--697},
  year    = {2026},
  doi     = {10.1107/S1600577526001943}
}

@article{yamaji2021,
  author  = {Yamaji, Youhei and Yoshida, Teppei and Fujimori, Atsushi and
             Imada, Masatoshi},
  title   = {Hidden self-energies as origin of cuprate superconductivity
             revealed by machine learning},
  journal = {Physical Review Research},
  volume  = {3},
  number  = {4},
  pages   = {043099},
  year    = {2021},
  doi     = {10.1103/PhysRevResearch.3.043099}
}

@article{yamaji2023erratum,
  author  = {Yamaji, Youhei and Yoshida, Teppei and Fujimori, Atsushi and Imada, Masatoshi},
  title   = {Erratum: Hidden self-energies as origin of cuprate superconductivity revealed by machine learning [{Phys}. {Rev}. {Research} {\bf 3}, 043099 (2021)]},
  journal = {Physical Review Research},
  volume  = {5},
  number  = {1},
  pages   = {019003},
  year    = {2023},
  doi     = {10.1103/PhysRevResearch.5.019003}
}

@article{petkovic2025machine,
  title={Machine learning in crystal growth: A review of methods, data, and applications},
  author = {Petkovic, M and Vieira, L and Dropka, N},
  journal={Progress in Crystal Growth and Characterization of Materials},
  volume={71},
  number={4},
  pages={100689},
  year={2025},
  publisher={Elsevier},
  doi={10.1016/j.pcrysgrow.2025.100689}
}

@article{yan2025large,
  title     = {Large scale and diverse two-dimensional flake segmentation dataset by general-purpose and labor-efficient annotation framework},
  author = {Yan, Su and Chen, Jun and Li, Xiao and Bai, Jinfan and Zhou, Rong and Zhang, Liping and Li, Weijun and Wu, Jiang-Bin and Tan, Ping-Heng and Ning, Xin},
  journal   = {npj 2D Materials and Applications},
  volume    = {9},
  number    = {1},
  pages     = {103},
  year      = {2025},
  publisher = {Nature Publishing Group UK London},
  doi       = {10.1038/s41699-025-00622-9}
}

@article{zhang2026,
  author  = {Zhang, Yu and Zhong, Yong and Tran, Nhat Huy and Li, Shuyi and
             Lee, Kyuho and Lee, Yonghun and Wang, Tiffany C. and
             Hwang, Harold Y. and Shen, Zhi-Xun and Jia, Chunjing},
  title   = {Machine learning reconstruction of high-dimensional electronic
             structure from angle-resolved photoemission spectroscopy},
  journal = {arXiv preprint},
  year    = {2026},
  eprint  = {2603.16725},
  archivePrefix = {arXiv},
  url     = {https://arxiv.org/abs/2603.16725}
}

@article{xarpes2026,
  author = {van Waas, Thomas P. and Berthod, Christophe and Berges, Jan and Marzari, Nicola and Dil, J. Hugo and Ponc{\'e}, Samuel},
  title   = {Extraction of the self energy and {Eliashberg} function from
             angle-resolved photoemission spectroscopy using the {xARPES}
             code},
  journal = {npj Computational Materials},
  volume  = {12},
  pages   = {172},
  year    = {2026},
  doi     = {10.1038/s41524-026-02026-9}
}

@article{deepprior2026,
  author = {Yokoyama, Yuichi and Yamagami, Kohei and Sumiya, Yuta and Shouno, Hayaru and Mizumaki, Masaichiro},
  title   = {Deep prior-based denoising for state-of-the-art scientific imaging and metrology},
  journal = {Machine Learning: Science and Technology},
  volume  = {7}, number = {3}, pages = {035048}, year = {2026},
  doi     = {10.1088/2632-2153/ae67cf}
}

@article{deephmagnetic2023,
  author = {Li, He and Tang, Zechen and Gong, Xiaoxun and Zou, Nianlong and Duan, Wenhui and Xu, Yong},
  title   = {Deep-learning electronic-structure calculation of magnetic superstructures},
  journal = {Nature Computational Science},
  volume  = {3}, pages = {321--327}, year = {2023},
  doi     = {10.1038/s43588-023-00424-3}
}

@article{deephhybrid2024,
  author = {Tang, Zechen and Li, He and Lin, Peize and Gong, Xiaoxun and Jin, Gan and He, Lixin and Jiang, Hong and Ren, Xinguo and Duan, Wenhui and Xu, Yong},
  title   = {A deep equivariant neural network approach for efficient hybrid density functional calculations},
  journal = {Nature Communications},
  volume  = {15}, pages = {8815}, year = {2024},
  doi     = {10.1038/s41467-024-53028-4}
}

@article{deephplanewave2024,
  author = {Gong, Xiaoxun and Louie, Steven G. and Duan, Wenhui and Xu, Yong},
  title   = {Generalizing deep learning electronic structure calculation to the plane-wave basis},
  journal = {Nature Computational Science},
  volume  = {4}, year = {2024},
  doi     = {10.1038/s43588-024-00701-9}
}

@article{knosgaard2022,
  author = {Kn{\o}sgaard, Nikolaj R{\o}rb{\ae}k and Thygesen, Kristian Sommer},
  title   = {Representing individual electronic states for machine learning {GW} band structures of 2D materials},
  journal = {Nature Communications},
  volume  = {13}, pages = {468}, year = {2022},
  doi     = {10.1038/s41467-022-28122-0}
}

@article{kohnshamvae2024,
  author = {Hou, Bowen and Wu, Jinyuan and Qiu, Diana Y.},
  title   = {Unsupervised representation learning of {Kohn--Sham} states and consequences for downstream predictions of many-body effects},
  journal = {Nature Communications},
  volume  = {15}, pages = {9481}, year = {2024},
  doi     = {10.1038/s41467-024-53748-7}
}

@article{slakonet2025,
  author = {Choudhary, Kamal},
  title   = {{SlaKoNet}: a unified {Slater--Koster} tight-binding framework using neural network infrastructure for the periodic table},
  journal = {The Journal of Physical Chemistry Letters},
  volume  = {16}, number = {43}, pages = {11109--11119}, year = {2025},
  doi     = {10.1021/acs.jpclett.5c02456}
}

@article{chen2021review,
  author = {Chen, Zhantao and Andrejevic, Nina and Drucker, Nathan C. and Nguyen, Thanh and Xian, R. Patrick and Smidt, Tess and Wang, Yao and Ernstorfer, Ralph and Tennant, D. Alan and Chan, Maria and Li, Mingda},
  title   = {Machine learning on neutron and x-ray scattering and spectroscopies},
  journal = {Chemical Physics Reviews},
  volume  = {2}, number = {3}, pages = {031301}, year = {2021},
  doi     = {10.1063/5.0049111}
}

@article{pustovit2024,
  author = {Pustovit, Yu. V. and Lytveniuk, Ye. P.},
  title   = {Neural-network-based methods for {ARPES} data processing ({Review} article)},
  journal = {Ukrainian Journal of Physics},
  volume  = {69},
  number  = {1},
  pages   = {53--65},
  year    = {2024},
  doi     = {10.15407/ujpe69.1.53}
}

@inproceedings{vincent2008,
  author = {Vincent, Pascal and Larochelle, Hugo and Bengio, Yoshua and Manzagol, Pierre-Antoine},
  title     = {Extracting and composing robust features with denoising autoencoders},
  booktitle = {Proceedings of the 25th International Conference on Machine Learning (ICML)},
  pages     = {1096--1103},
  year      = {2008},
  doi       = {10.1145/1390156.1390294}
}

@inproceedings{he2020moco,
  author    = {He, Kaiming and Fan, Haoqi and Wu, Yuxin and Xie, Saining and Girshick, Ross},
  title     = {Momentum Contrast for Unsupervised Visual Representation Learning},
  booktitle = {Proceedings of the {IEEE}/{CVF} Conference on Computer Vision and Pattern Recognition ({CVPR})},
  pages     = {9729--9738},
  year      = {2020},
  doi       = {10.1109/CVPR42600.2020.00975}
}

@inproceedings{caron2020swav,
  author    = {Caron, Mathilde and Misra, Ishan and Mairal, Julien and Goyal, Priya and Bojanowski, Piotr and Joulin, Armand},
  title     = {Unsupervised Learning of Visual Features by Contrasting Cluster Assignments},
  booktitle = {Advances in Neural Information Processing Systems},
  volume    = {33},
  pages     = {9912--9924},
  year      = {2020}
}

@inproceedings{grill2020byol,
  author    = {Grill, Jean-Bastien and Strub, Florian and Altch{\'e}, Florent and Tallec, Corentin and Richemond, Pierre H. and Buchatskaya, Elena and Doersch, Carl and Pires, Bernardo Avila and Guo, Zhaohan Daniel and Azar, Mohammad Gheshlaghi and Piot, Bilal and Kavukcuoglu, Koray and Munos, R{\'e}mi and Valko, Michal},
  title     = {Bootstrap Your Own Latent: A New Approach to Self-Supervised Learning},
  booktitle = {Advances in Neural Information Processing Systems},
  volume    = {33},
  pages     = {21271--21284},
  year      = {2020}
}

@inproceedings{caron2021dino,
  author    = {Caron, Mathilde and Touvron, Hugo and Misra, Ishan and J{\'e}gou, Herv{\'e} and Mairal, Julien and Bojanowski, Piotr and Joulin, Armand},
  title     = {Emerging Properties in Self-Supervised Vision Transformers},
  booktitle = {Proceedings of the {IEEE}/{CVF} International Conference on Computer Vision ({ICCV})},
  pages     = {9650--9660},
  year      = {2021}
}

@article{raissi2019,
  author = {Raissi, Maziar and Perdikaris, Paris and Karniadakis, George E.},
  title   = {Physics-informed neural networks: a deep learning framework for solving forward and inverse problems involving nonlinear partial differential equations},
  journal = {Journal of Computational Physics},
  volume  = {378}, pages = {686--707}, year = {2019},
  doi     = {10.1016/j.jcp.2018.10.045}
}

@article{majchrzak2025,
  author = {Majchrzak, Paulina and Sanders, Charlotte and Zhang, Yu and Kuibarov, Andrii and Suvorov, Oleksandr and Springate, Emma and Kovalchuk, Iryna and Aswartham, Saicharan and Shipunov, Grigory and B{\"u}chner, Bernd and Yaresko, Alexander and Borisenko, Sergey and Hofmann, Philip},
  title   = {Machine-learning approach to understanding ultrafast carrier dynamics in the three-dimensional {Brillouin} zone of {PtBi$_2$}},
  journal = {Physical Review Research},
  volume  = {7}, number = {1}, pages = {013025}, year = {2025},
  doi     = {10.1103/PhysRevResearch.7.013025}
}

@article{denby1988,
  author = {Denby, Bruce},
  title   = {Neural networks and cellular automata in experimental high energy physics},
  journal = {Computer Physics Communications},
  volume  = {49},
  number  = {3},
  pages   = {429--448},
  year    = {1988},
  doi     = {10.1016/0010-4655(88)90004-5}
}

@phdthesis{staab2024,
  title  = {Using micro{ARPES} and Machine Learning to Accelerate Experiments and Elucidate the Local Electronic and Chemical Structure of Quantum Materials},
  author = {Staab, Michael C.},
  school = {University of California, Davis},
  year   = {2024},
  type   = {{Ph.D.} thesis},
  url    = {https://escholarship.org/uc/item/1207d032},
  note   = {ProQuest ID: Staab\_ucdavis\_0029D\_23635, Merritt ID: ark:/13030/m5cp8mx8}
}

@misc{staab2026,
  title = {Electronic and Chemical Phase Identification in Photoemission Experiments Using Unsupervised Machine Learning},
  author = {Staab, Matthew and Pandur, Joseph and Rotenberg, Eli and Jozwiak, Chris and Bostwick, Aaron and Vishik, Inna},
  year = 2026,
  month = aug,
  number = {arXiv:2608.19503},
  eprint = {2608.19503},
  primaryclass = {cond-mat.mes-hall},
  publisher = {arXiv},
  doi = {10.48550/arXiv.2608.19503},
  urldate = {2026-09-11},
  archiveprefix = {arXiv},

}

@article{lehoux2026benchmarking,
  author        = {Le Houx, James},
  title         = {Benchmarking Autonomy in Scientific Experiments: A Hierarchical Taxonomy for Autonomous Large-Scale Facilities},
  journal       = {arXiv preprint},
  year          = {2026},
  eprint        = {2601.06978},
  archivePrefix = {arXiv},
  primaryClass  = {physics.ins-det},
  url           = {https://arxiv.org/abs/2601.06978},
  note          = {Introduces the {BASE} Scale (Benchmarking Autonomy in Scientific Experiments)}
}

@techreport{saej3016_2021,
  author = {{SAE International}},
  title       = {Taxonomy and Definitions for Terms Related to Driving Automation Systems for On-Road Motor Vehicles},
  type        = {Standard},
  institution = {SAE International},
  number      = {J3016},
  year        = {2021},
  url         = {https://www.sae.org/standards/j3016_202104}
}

@inproceedings{he2020wave,
  title={Wave optics simulations for new soft x-ray beamlines at NSLS-II},
  author = {He, An and Chubar, Oleg and Dvorak, Joseph and Hu, Wen and Hulbert, Steven and Walter, Andrew and Wilkins, Stuart and Salort, Coral},
  booktitle={Advances in Computational Methods for X-Ray Optics V},
  editor={Sawhney, Kawal and Chubar, Oleg},
  year={2020},
  volume={11493},
  pages={114930H},
  address={Bellingham, WA},
  publisher={SPIE},
  doi={10.1117/12.2567423}
}

@article{Mannix2022,
  author = {Mannix, Andrew J. and Ye, Andrew and Sung, Suk Hyun and Ray, Ariana and Mujid, Fauzia and Park, Chibeom and Lee, Myungjae and Kang, Jong Hoon and Shreiner, Robert and High, Alexander A. and Muller, David A. and Hovden, Robert and Park, Jiwoong},
  title   = {Robotic four-dimensional pixel assembly of van der Waals solids},
  journal = {Nature Nanotechnology},
  year    = {2022},
  volume  = {17},
  number  = {4},
  pages   = {361--366},
  doi     = {10.1038/s41565-021-01061-5}
}

@misc{Park2024,
  author = {Park, Suji and Yager, Kevin},
  title = {Roll-to-roll mechanized exfoliator and automatic 2D materials transfer and layering system},
  year = {2024},
  month = {November 14},
  note = {US Patent App. 18/291,428, Publication No. US 2024/0375960 A1},
  url = {https://patentimages.storage.googleapis.com/14/b0/13/d19a797f716917/US20240375960A1.pdf}
}

@article{stach2021,
  title = {Autonomous Experimentation Systems for Materials Development: {{A}} Community Perspective},
  author = {Stach, Eric and DeCost, Brian and Kusne, A Gilad and {Hattrick-Simpers}, Jason and Brown, Keith A and Reyes, Kristofer G and Schrier, Joshua and Billinge, Simon and Buonassisi, Tonio and Foster, Ian and Gomes, Carla P and Gregoire, John M and Mehta, Apurva and Montoya, Joseph and Olivetti, Elsa and Park, Chiwoo and Rotenberg, Eli and Saikin, Semion K and Smullin, Sylvia and Stanev, Valentin and Maruyama, Benji},
  year = 2021,
  month = sep,
  journal = {MATTER},
  volume = {4},
  number = {9},
  pages = {2702--2726},
  publisher = {ELSEVIER},
  address = {RADARWEG 29, 1043 NX AMSTERDAM, NETHERLANDS},
  issn = {2590-2393},
  doi = {10.1016/j.matt.2021.06.036}
}

@misc{anl_rapid200,
  author       = {{Argonne National Laboratory}},
  title        = {{RAPID-200}: Autonomous Lab for Materials and Chemistry},
  year         = {2025},
  howpublished = {\url{https://www.anl.gov/autonomous-discovery/rapid200-autonomous-lab-for-materials-and-chemistry}},
  note         = {Accessed 2026}
}

@misc{royce_uhv2d,
  author       = {{Henry Royce Institute}},
  title        = {{UHV} {2D} Materials and Devices Assembly},
  year         = {2025},
  howpublished = {\url{https://www.royce.ac.uk/technology-platforms/uhv-2d-materials/}},
  note         = {Platform lead: Roman Gorbachev, University of Manchester. Accessed 2026}
}

@misc{han_erlab,
  author    = {Higashihira Han, Kimoon},
  title     = {erlab: A Complete Python Workflow for {ARPES} Experiments},
  year      = {2026},
  version   = {v3.27.3},
  publisher = {Zenodo},
  doi       = {10.5281/zenodo.16809480}
}

@misc{ekahana_arpesskill,
  author    = {Ekahana, Sandy Adhitia},
  title     = {{\texttt{ARPESskill}}: General {LLM} Agent Skill for {ARPES} Analysis},
  year      = {2026},
  publisher = {GitHub},
  url       = {https://github.com/fawkesdx/ARPESskill}
}

@misc{doe_isaac_pathfinder,
  author    = {{U.S. Department of Energy, Basic Energy Sciences}},
  title     = {Integrated Scientific Agentic {AI} for Catalysis ({ISAAC}): Basic Energy Sciences {AI} Pathfinder Program},
  year      = {2024},
  publisher = {SLAC National Accelerator Laboratory},
  url       = {https://isaac.slac.stanford.edu/}
}
\end{document}